\documentclass{article}

\usepackage[american]{babel}
\usepackage{csquotes}
\usepackage[style=authoryear-comp,citetracker=true,maxbibnames=99,maxcitenames=2,giveninits=true,backend=bibtex]{biblatex}
\DeclareNameAlias{sortname}{last-first}
\AtBeginBibliography{%
}
\renewbibmacro{in:}{}
\DeclareFieldFormat[article]{title}{#1}
\DeclareFieldFormat{pages}{#1}
\renewbibmacro*{volume+number+eid}{%
  \printfield{volume}%
  \setunit*{\addnbthinspace}
  \printfield{number}%
  \setunit{\addcomma\space}%
  \printfield{eid}}
\DeclareFieldFormat[article]{number}{\mkbibparens{#1}}
\DeclareFieldFormat[article]{volume}{\textit{#1}}

\DeclareFieldFormat{sentencecase}{\MakeSentenceCase{#1}}

\renewbibmacro*{title}{%
  \ifthenelse{\iffieldundef{title}\AND\iffieldundef{subtitle}}
    {}
    {\ifthenelse{\ifentrytype{article}\OR\ifentrytype{inbook}%
      \OR\ifentrytype{incollection}\OR\ifentrytype{inproceedings}%
      \OR\ifentrytype{inreference}}
      {\printtext[title]{%
        \printfield[sentencecase]{title}%
        \setunit{\subtitlepunct}%
        \printfield[sentencecase]{subtitle}}}%
      {\printtext[title]{%
        \printfield[titlecase]{title}%
        \setunit{\subtitlepunct}%
        \printfield[titlecase]{subtitle}}}%
     \newunit}%
  \printfield{titleaddon}}

\AtEveryCitekey{\ifciteseen{}{\defcounter{maxnames}{99}}}

\usepackage{amsmath}
\usepackage{amsfonts}
\usepackage{amssymb}
\usepackage{amsthm}
\usepackage{chngcntr}
\usepackage{graphicx}
\usepackage{subcaption}
\usepackage{fancyhdr}
\usepackage{tabularx}
\usepackage{geometry}
\usepackage{setspace}
\usepackage[right]{eurosym}
\usepackage[printonlyused]{acronym}
\usepackage{floatflt}
\usepackage[capposition=top]{floatrow}
\usepackage{colortbl}
\usepackage{paralist}
\usepackage{array}
\usepackage{titlesec}
\usepackage{parskip}
\usepackage[right]{eurosym}
\usepackage{longtable}
\usepackage[colorlinks=true,linkcolor=black,urlcolor=blue,citecolor=black,pdfpagelabels=true]{hyperref}
\usepackage{mathtools}
\usepackage{booktabs,caption}
\usepackage[flushleft]{threeparttable}
\usepackage{hyperref}
\usepackage{dsfont}
\usepackage{enumitem}
\usepackage[toc,page]{appendix}
\usepackage{rotating}
\usepackage{bbold}
\usepackage{scrextend}
\usepackage{xcolor}
\usepackage{algorithm2e}

\usepackage{listings}
\makeatletter
\def\l@lstlisting#1#2{\@dottedtocline{1}{0em}{1em}{\hspace{1,5em} Lst. #1}{#2}}
\makeatother

\newcommand\posscite[1]{\citeauthor{#1}'s (\citeyear*{#1})}

\numberwithin{equation}{section}

\theoremstyle{plain}
\newtheorem{assumption}{Assumption}
\newtheorem{assumptionprime}{Assumption}

\newtheorem{theorem}{Theorem}

\newtheorem{definition}{Definition}
\begin{document}

\renewcommand*{\thefootnote}{\fnsymbol{footnote}}
\thispagestyle{empty}
\begin{center}
	\Large
	\textbf{Nonparametric estimation of causal heterogeneity under high-dimensional confounding}\\~\\
\vspace*{2cm}
\begin{large}
	Michael Zimmert\footnote[2]{Email: michael.zimmert@unisg.ch} and Michael Lechner\footnote[3]{Michael Lechner is also affiliated with CEPR, London, CESIfo, Munich, IAB, Nuremberg, IZA, Bonn, and RWI, Essen.\\
	Email: michael.lechner@unisg.ch, www.michael-lechner.eu}\\~\\
	SEW-HSG\\
Swiss Institute for Empirical Economic Research\\
University of St.Gallen, Switzerland\\
	
	\end{large}
\vspace*{1cm}
\begin{abstract}
This paper considers the practically important case of nonparametrically estimating heterogeneous average treatment effects that vary with a limited number of discrete and continuous covariates in a selection-on-observables framework where the number of possible confounders is very large. We propose a two-step estimator for which the first step is estimated by machine learning. We show that this estimator has desirable statistical properties like consistency, asymptotic normality and rate double robustness. In particular, we derive the coupled convergence conditions between the nonparametric and the machine learning steps. We also show that estimating population average treatment effects by averaging the estimated heterogeneous effects is semi-parametrically efficient. The new estimator is an empirical example of the effects of mothers’ smoking during pregnancy on the resulting birth weight.
\end{abstract}	
	
	\vspace*{5.5cm}
\end{center}
\vspace*{\fill}
\textbf{JEL classification:} C14, C21\\
\textbf{Keywords:} causal machine learning, effect heterogeneity, group average treatment effects, semiparametric efficiency, ensemble learning\\
\pagebreak

\renewcommand*{\thefootnote}{\arabic{footnote}}
\setcounter{footnote}{0}
\setcounter{page}{1}
\doublespacing
\allowdisplaybreaks
\section{Introduction}\label{sec:introduction}
Recently, new machine learning based estimators showed immense potential to systematically uncovering causal effect heterogeneity so that there is now a rapidly growing literature on this topic (e.g., see the overviews in \cite{Athey_Imbens_2017,Athey_Imbens_2019} and \cite{Knaus_Lechner_Strittmatter_2018}). In the context of heterogeneity, the respective aggregation levels for which the heterogeneity is estimated is playing an important role. Most papers of this literature focus on a selection-on-observable framework and investigate estimators for the heterogeneity at the lowest aggregation level to uncover possible heterogeneities to the largest extent possible. While this finest level of causal granularity is obviously of interest, \textcite{Chernozhukov_FernandezVal_Luo_2018} and \textcite{Lechner_2018} argue to analyse heterogeneity at higher levels, so called `Group Average Treatment Effects' (GATEs). Such aggregates can be estimated more precisely, may be far more easily interpretable by researchers in substantive terms, and are more useful for decision makers. In particular, some subgroup heterogeneities are of limited value per se because it is hard to justify a decision or policy based on certain characteristics (race, gender etc.). Therefore, decision-makers are often only interested in effect heterogeneities based on a rather small subset of available covariates. This paper suggests an approach that is based on statistical-learning assisted estimation of the GATEs for the various discrete and continuous variables of interest, and subsequent non-parametric aggregation of the GATEs to obtain `Average Treatment Effects' (ATEs).\footnote{\textcite{Lechner_2018} also proposed this aggregation idea. However, that paper considered only a version of a Causal Forest while here we are in principle agnostic with respect to the machine learning method used. Furthermore, it considered only GATEs based on discrete variables and thus GATEs were obtained as unweighted within-cell means.}\\
More technically speaking, in effect heterogeneity analysis covariates do not (only) serve the purpose of making identifying assumptions credible. They become part of the outcome analysis by discriminating different subgroups of units for which the effect is of interest. Further, whenever new observations enter the sample the covariate realizations could be used to predict a causal effect. The set of covariates to be included in the statistical model to explore effect heterogeneity is therefore not a statistical but rather a substantive decision.\\
The estimation of subgroup specific effects is a tedious task when there is confounding. In such settings, causal effects are typically only identified if the researcher includes the confounding covariates in the statistical model as well. Hence, the identifying assumptions dictate the inclusion of the set of covariates required. In empirical research based on selection-on-observables, the credibility of causal effects estimation often depends on a very large set of possible covariates with very many possible functional forms. Qualitatively assessing which covariates should ultimately enter the model in which specific form in a non-systematic fashion is prone to be flawed.\\
There are currently several suggestions to estimate heterogeneous effects when there is confounding. The general concept of estimating causal effects conditional on covariates\footnote{We avoid the imprecise term `Conditional Average Treatment Effect' because it is unclear which conditioning set is actually meant.} already dates back to \textcite{Hahn_1998}. He suggested estimating a nonparametric outcome regression on the set of covariates that needs to be controlled for. Averaging over the conditional means leads to estimators of ATE that attain the semiparametric efficiency bound. In practice, however, nonparametric regression with many covariates is hardly feasible because the convergence rate of nonparametric methods exponentially decreases with the number of covariates included. Recently, \textcite{Wager_Athey_2018} follow the same ideas as in \textcite{Hahn_1998} but use Causal Forests instead of standard nonparametric regression. \textcite{Athey_Tibshirani_Wager_2019} and \textcite{Lechner_2018} modify the Random Forest algorithm to better adjust for confounding and improve precision. Outcome-based models adjust for confounding and infer heterogeneous  effects in a single estimation step. Therefore, in all of these approaches inference for effect heterogeneity relies on a dimension of the covariate space that is fixed. Given the previous discussion, this might be a very strong assumption.\\
In this paper we follow an alternative approach in the literature. The two distinct roles of the covariates -- adjusting for confounding and estimating heterogeneous effects -- are explicitly reflected in a two-step estimation procedure for the GATEs. This idea is conceptionally not new in the literature.\footnote{A few days before this work appeared first on arXiv, \textcite{Fan_Hsu_Lieli_Zhang_2019} published their independent work on arXiv (up to that moment unknown to us) that uses similar ideas about aggregation and machine learning.} In the context of difference-in-differences estimation, \textcite{Abadie_2005} shows that propensity scores weighted outcomes can be used as a dependent variable in a second stage regression on the covariates that are of interest for heterogeneous effects. \textcite{Abrevaya_Hsu_Lieli_2015} use a similar idea in the standard selection-on-observables setting. They provide inferential results for nonparametric and parametric propensity score first stages with nonparametric second stages. In line with other results in the literature on average effects (\cite{Hirano_Imbens_Ridder_2003}, \cite{Robins_Rotnitzky_Zhao_1994}, \cite{Lunceford_Davidian_2004}), they show that the variance for Inverse Probability Weighting (IPW) estimators can be substantially decreased when the propensity score is estimated nonparametrically. Since their second stage also relies on nonparametric regression, the validity of their asymptotic results requires jointly choosing two kernel bandwidths which have to be in a rather small feasible interval. \textcite{Lee_Okui_Whang_2016} augment the model of \textcite{Abrevaya_Hsu_Lieli_2015} by including outcome projections (Augmented IPW, AIPW) and show that when both the propensity score and the outcome projections are estimated parametrically, one can treat the nuisance parameters as if they were known. Their asymptotic results with parametric first stages are then equivalent\footnote{While \textcite{Abrevaya_Hsu_Lieli_2015} use local constant nonparametric regression, \textcite{Lee_Okui_Whang_2016} show their results with local linear nonparametric regression.} to those of \textcite{Abrevaya_Hsu_Lieli_2015} with nonparametric propensity score estimation. Still the parametric nuisance models used could lead to substantial misspecification bias if the functional form does not coincide with the unknown data generating process (DGP). Moreover, if there are more potential regressors than observations even parametric estimators collapse.\\
In their contributions to ATE estimation \textcite{Belloni_Chernozhukov_Hansen_2014} and \textcite{Chernozhukov_Chetverikov_Demirer_Duflo_Hansen_Newey_2017} show how AIPW type estimators can be adapted to settings where the dimension of the relevant confounders grows with the sample size. They use various machine learning methods to estimate the propensity score and the outcome projections. Recently, \textcite{Chernozhukov_Semenova_2017} use their framework to estimate effect heterogeneity based on linear models. They provide conditions under which their second-stage linear model can become increasingly flexible.\\
Postulating nonparametric second stages, we do not assume any specific functional form of the GATE. We contribute to the literature on GATE estimation by allowing for flexible functional forms as well as a high-dimensional\footnote{In general the term `high-dimensional' refers to the fact that the dimension of the model can grow with the sample size. We will provide specific rate conditions in the main part of the paper.} confounder space. This enables our proposed estimator to be robust against functional form misspecification and to remain consistent even if the number of covariates relative to the sample size is large. In particular, we provide a generic statistical framework such that the convergence rate requirements of the first stage nuisance estimation are coupled with the kernel bandwidth second stage nonparametric convergence.\\
Additionally, we link our identification and estimation result to semiparametric efficiency theory by providing a new estimator for the ATE that can be estimated as a by-product of the GATEs. The estimator aggregates over all point estimates of the GATEs. We show that under certain convergence conditions for the kernel bandwidth, asymptotically it hits the variance lower bound of the semiparametric estimation problem. We therefore also contribute to the small literature on three-step semiparametric ATE estimation. Specifically, \textcite{Hahn_Ridder_2013} (for an alternative theoretical development see also \cite{Mammen_Rothe_Schienle_2012}) investigate a related set-up showing that nonparametric regression on an estimated propensity score can lead to efficient estimation of ATE. To the best of our knowledge, this paper is, however, the first that analyses the asymptotic properties of averaging a transformed outcome projection instead of the outcome projection on the propensity score. Like propensity score matching, our three-step estimator might have better finite-sample properties than two-step estimators that share the same first-order asymptotic properties (\cite{Robins_Rotnitzky_Zhao_1994}, \cite{Hirano_Imbens_Ridder_2003}, \cite{Chernozhukov_Chetverikov_Demirer_Duflo_Hansen_Newey_2017}) because the propensity score weights are subject to an additional smoothing step. Unlike propensity score matching, IPW or \posscite{Hahn_1998} estimator, the proposed ATE estimator remains feasible when the dimension of the confounders entering the model is high.\\
After providing some more information on the theoretical background in Section \ref{sec:background}, we present the details of our main asymptotic results in Section \ref{sec:results}. An empirical example in Section \ref{sec:example} compares different alternative estimators for GATE and ATE and illustrates the applicability and usefulness of the new methods. The last section concludes. The formal proofs of our theorems as well as some details on the empirical implementation are relegated to the Appendix.
\section{Methodology}\label{sec:background}
\subsection{Notation}
Suppose that we observe an independent and identically distributed random sample $\{w_i\}_{i=1}^N$ with sample size $N$ where $w_i=(y_i,d_i,x_i,z_i)$. Denote with uppercase letters a variable and with lowercase letters its realizations. Then $Y$ is the outcome variable and $D$ is the binary treatment of interest. To describe causal effects, we use \posscite{Rubin_1974} potential outcome notation such that $Y^d$ is the outcome that would have been observed under treatment $D=d$. Further, $X$ is a matrix of observed covariates with support $\mathcal{X}$ and $Z\subseteq X$ as a set of predefined variables where the researcher is interested in effect heterogeneity with support $\mathcal{Z}$. Also let $X\in\mathbb{R}^{\dim{X}}$ and $Z\in\mathbb{R}^{\dim{Z}}$ and denote $\lambda_X=\dim{X}$ and $\lambda_Z=\dim{Z}$. Potentially we have that $\lambda_X\rightarrow\infty$ when $N\rightarrow\infty$ whereas $\lambda_Z$ is fixed.\footnote{The concrete growth rates of $\lambda_X$ in relation to $N$ will be discussed in Section \ref{sec:results}.} Hence, we explicitly allow for models where the dimension of $X$ is high-dimensional but the dimension of the subset of covariates that is of interest for the heterogeneity analysis does not grow with the sample size. We remain agnostic about the underlying cumulative distribution from which the sample of $W=(Y,D,X,Z)$ is drawn $F=F(W)$ and just assume that it exists with density $f=f(W)$.
\subsection{Semiparametric efficiency theory}\label{subsec:semiparametric}
The main parameter of interest in this study is the GATE defined as
\begin{align*}
\tau(z)=\mathbb{E}\left[Y^1-Y^0|Z=z\right].
\end{align*}
Since we want to avoid usually unrealistic parametric assumptions on the underlying DGP, we allow for a flexible function $\psi(W,\cdot)$ such that GATE is identified as
\begin{align*}
\tau(z)=\mathbb{E}\left[\psi(W,\cdot)|Z=z\right]
\end{align*}
using the variables observed or functions of them. Many possible transformations of the outcome exists (see the references mentioned in the Introduction). It is, however, a priori unclear which one is a `good' transformation in the sense that it achieves the variance lower bound for the problem. Hence, ideally our exposition would start by deriving the semiparametric efficiency bound for the problem at hand and then use the moment condition implied as an estimand for GATE. However, since parameters using `last stage' nonparametric projections are not pathwise differentiable, standard semiparametric efficiency bounds cannot be derived following established theory (e.g. \cite{Bickel_Klaassen_Ritov_Wellner_1993}, \cite{Newey_1994}, \cite{Hahn_1998}, \cite{Tsiatis_2006}). This was also noted for different problems in \textcite{Rubin_vanderLaan_2007} and \textcite{Kennedy_Ma_McHugh_Small_2017}. We follow their approaches. Instead of directly relying on an efficiency result for GATE, we use that
\begin{align*}
\text{ATE}=\theta=\mathbb{E}\left[\mathbb{E}\left[\psi(W,\cdot)|Z=z\right]\right]
\end{align*}
implying \posscite{Hahn_1998} efficient score function for ATE
\begin{align*}
\psi(W,p,m_0,m_1)=\frac{D(Y-m_1(X))}{p(X)}-\frac{(1-D)(Y-m_0(X))}{1-p(X)}+m_1(X)-m_0(X)
\end{align*}
where $p(X)=\mathbb{E}\left[D|X\right]$ denotes the propensity score and $m_d(X)=\mathbb{E}\left[Y|X,D=d\right]$ for $d\in\{0,1\}$ denotes the conditional expectations of the outcome in the treatment-specific subpopulations.
\subsection{Parameter identification}
For identification of GATE and ATE we make the following assumptions.\\
\begin{assumption}[Conditional independence]\label{ass:cia}
\begin{align*}
Y^0,Y^1\perp D|X=x \quad \forall x\in\mathcal{X}\\
\end{align*}
\end{assumption}
\begin{assumption}[Stable Unit Treatment Value Assumption (SUTVA)]
\begin{align*}
Y=DY^1+(1-D)Y^0\\
\end{align*}
\end{assumption}
\begin{assumption}[Exogeneity of confounders]
\begin{align*}
X^1=X^0\\
\end{align*}
\end{assumption}
\begin{assumption}[Common support]\label{ass:cs}
\begin{align*}
c<p(X)<1-c
\end{align*}
for some small positive constant $c$.\\
\end{assumption}
Assuming that appropriate moments exist, then for GATE we have
\begin{align*}
\tau(z)=&\mathbb{E}\left[\mathbb{E}\left(Y^1-Y^0|X\right)|Z=z\right]\\
=&\mathbb{E}\left[\frac{D(Y-m_1(X))}{p(X)}-\frac{(1-D)(Y-m_0(X))}{1-p(X)}+m_1(X)-m_0(X)\Big|Z=z\right]\\
=&\mathbb{E}\left[\frac{DY}{p(X)}-\frac{(1-D)Y}{1-p(X)}\Big|Z=z\right]\\
=&\mathbb{E}\left[m_1(X)-m_0(X)|Z=z\right].
\end{align*}
The exposition shows that IPW and outcome based estimands are embedded in the estimand based on $\psi(W,p,m_0,m_1)$. Finally, by noticing that
\begin{align*}
\theta=\mathbb{E}\left[\tau(Z)\right]
\end{align*}
identification of ATE trivially follows from these considerations.
\section{Main results}\label{sec:results}
\subsection{GATE estimation}
\subsubsection{Proposed estimator}
The identification results from the preceding section suggest a two-step estimation strategy. The details of our proposed estimator are described in Procedure 1.
\begin{figure}[h]
\centering
\onehalfspacing
\fbox{\begin{minipage}{0.95\textwidth}
\textbf{Procedure 1.} GATE estimation\\~\\
Introduce the subsample index $l=1,...,L$ and denote the corresponding information set by $\mathcal{I}_l$ as well as its complement by $\mathcal{I}_l^C$.
\begin{enumerate}
\item Randomly split the sample in equally sized subsamples $1,...,L$.
\item \textbf{for} $l=1$ \textbf{to} $L$ \textbf{do}
\begin{itemize}
\item[] Estimate the propensity score $p(x)$ and the outcome projections $m_0(x)$ and $m_1(x)$ in the sample with $\mathcal{I}_l^C$ using any suitable machine learning method or an ensemble of them.
\item[] Predict $\hat{p}(x)$, $\hat{m}_0(x)$ and $\hat{m}_1(x)$ in the sample with $\mathcal{I}_l$.
\end{itemize}
\textbf{end}
\item Denote $\hat{p}=\hat{p}_{l=1,...,L}$, $\hat{m}_{0}=\hat{m}_{0,l=1,...,L}$ and $\hat{m}_{1}=\hat{m}_{1,l=1,...,L}$. Then construct the vector with elements $\hat{\psi}=\psi(W_i,\hat{p},\hat{m}_{0},\hat{m}_{1})$ for $i=1,...,N$ and estimate GATE as 
\begin{align*}
\hat{\tau}(z)=\sum_{i=1}^N\frac{K\left(\frac{z_i-z}{h}\right)\psi(W_i,\hat{p},\hat{m}_0,\hat{m}_1)}{\sum_{i=1}^NK\left(\frac{z_i-z}{h}\right)}
\end{align*}
where $K=K(u)$ is some kernel function that depends on a bandwidth $h$.
\end{enumerate}
\end{minipage}}
\end{figure}
In a first step a sample plug-in versions of $\psi(W,p,m_0,m_1)$ can be obtained by estimating the nuisance parameters. In a second step the $\psi$-vector can be projected on $Z$. Our goal is to estimate both stages as flexible as possible and to avoid parametric assumptions. Further, our estimator can cope with settings where $\lambda_X$ is very large which precludes classical nonparametric and parametric methods to estimate the first stage nuisances $p(x)$, $m_0(x)$ and $m_1(x)$. However, we can use a large class of supervised machine learning algorithms that have been shown to be very effective predictors for such types of tasks. Following the suggestions of \textcite{Chernozhukov_Chetverikov_Demirer_Duflo_Hansen_Newey_2017} we apply a cross-fitting algorithm for the nuisance parameter estimation step in order to guarantee that the resulting estimator of $\psi(W,p,m_0,m_1)$ consists of independent observations. The requirements for the second stage estimation step are more sophisticated as this estimator should allow for valid inference. To estimate GATE flexibly, we apply nonparametric local constant regression in the second step.
\subsubsection{Asymptotic results}
We now investigate the theoretical properties of our proposed estimation procedure. To ease the notational burden, we start with some definitions.\\
\begin{definition}[Norms] Denote by $\lVert g(X)\rVert_p$ the $L_p$ norm of the generic function $g(\cdot)$. Further denote the supremum norm by $\sup_{X\in\mathcal{X}}\lvert g(X)\rvert=\lVert g(X)\rVert_{\infty}$.\\  
\end{definition}
\begin{definition}[Rates]\label{def:rates} The nuisance parameter first stage estimates $\hat{p}$, $\hat{m}_0$ and $\hat{m}_1$ obtained by the sample splitting procedure described above belong to the realization sets $\mathcal{P}$, $\mathcal{M}_0$ and $\mathcal{M}_1$ with probability $1-o(1)$. For any realization $p^*$, $m_0^*$ and $m_1^*$ in the sets define the rates
\begin{align*}
\epsilon_{m_d}&=\sup_{m_d^*\in\mathcal{M}_d}\lVert m_d^*(X)-m_d(X)\rVert_2\\
\epsilon_{p}&=\sup_{p^*\in\mathcal{P}}\lVert p^*(X)-p(X)\rVert_2\\
\epsilon_{\max}&=\max\{\epsilon_{m_0},\epsilon_{m_1},\epsilon_p\}.\\
\end{align*}
\end{definition}
\begin{definition}[Scaling factor]
For any function $g$ define a scaling parameter $\delta_g$ that determines $g=O\left(N^{-\delta_g}\right)$.\\
\end{definition}
We then make the following standard assumptions on the kernel regression step (see for example \cite{Pagan_Ullah_1999}).\\
\begin{assumption}[Kernel regression]\label{ass:kernel}~
\begin{enumerate}
\item $Z=z$ is a point in the interior of the support $\mathcal{Z}$.
\item The density function estimator is uniformly bounded away from zero such that $\inf_{z\in\mathcal{Z}}\hat{f}(z)\geq C$ where $C>0$ is a generic constant.
\item The Kernel function $K(u)$ is $r$ times continuously differentiable, symmetric and of order $r$ in the sense $\int u^{r-1}K(u)du=0$ and $\int u^{r}K(u)du=O(1)$ for $r\in\mathbb{N}$.
\item $f(z)$ and $\mathbb{E}\left(\psi(W,p,m_0,m_1)|Z=z\right)$ are $r$ times continuously differentiable.
\item Further the Kernel function satisfies (i) $\int K(u)du=1$, (ii) $\int \lvert K(u)\rvert^{2+C} du=O(1)$ for any $C>0$, (iii) $\lvert u\rvert\lvert K(u)\rvert\rightarrow 0$ as $\lvert u \rvert\rightarrow\infty$, (iv) $\left\lVert K(u)\right\rVert_{\infty}=O(1)$ and (v) $\int K^2(u)du=O(1)$.\\
\end{enumerate}
\end{assumption}
Assumption \ref{ass:kernel} comprises the standard nonparametric local constant regression assumptions allowing for multiple covariates and higher-order kernels. For illustrative purposes multivariate regression results are derived assuming the same bandwidth for every regressor. Further, we have to make boundedness assumptions on the second moment of the sample error of the outcome model and on the nuisance prediction errors.\\
\begin{assumption}[Boundedness of conditional variances]\label{ass:error}~
The conditional variances of the outcome models are bounded such that they obey
\begin{align*}
\mathbb{E}\left[\left(DY-m_1(X)\right)^2|X\right]=O(1)\quad \text{and} \quad \mathbb{E}\left[\left((1-D)Y-m_0(X)\right)^2|X\right]=O(1).\\
\end{align*}
\end{assumption}
\begin{assumption}[Boundedness of convergence rates]\label{ass:superror}~
The nuisance parameter prediction errors are bounded such that they obey
\begin{align*}
\sup_{m_d^*\in\mathcal{M}_d}\lVert m_d^*(X)-m_d(X)\rVert_\infty=O(1)\quad \text{for $d\in\{0,1\}$ and} \quad \sup_{p^*\in\mathcal{P}}\lVert p^*(X)-p(X)\rVert_\infty=O(1).\\
\end{align*}
\end{assumption}
Additionally, the convergence rates of our first stage nuisance parameter prediction and the second stage nonparametric regression are assumed to be as follows:\\
\begin{assumption}[Coupled convergence (GATE)]\label{ass:cate}~
The bandwidth $h$ and the sample size $N$ jointly converge such that
\begin{itemize}
\item[(i)] $h=o(1)$, $Nh^{\lambda_Z}\rightarrow\infty$ as $N\rightarrow\infty$ and
\item[(ii)] $N^{\frac{1}{2}}h^{\frac{1}{2}\lambda_Z}h^r=o(1)$.
\end{itemize}
Further, $N$ and $h$ satisfy the joint convergence conditions with the nuisance parameter convergence rates
\begin{itemize}
\item[(iii)] $h^{-\frac{1}{2}\lambda_Z}\epsilon_{\max}=o(1)$
\item[(iv)] $N^{\frac{1}{2}}h^{-\frac{1}{2}\lambda_Z}\epsilon_{m_0}\epsilon_p+N^{\frac{1}{2}}h^{-\frac{1}{2}\lambda_Z}\epsilon_{m_1}\epsilon_p=o(1)$.\\
\end{itemize}
\end{assumption}
Assumption \ref{ass:cate} comprises the coupled convergence rate assumptions that are at the centre of our theoretical results. Conditions (i) and (ii) quantify how the bandwidth has to converge to zero in relation to the sample size $N$ and the number of regressors $\lambda_Z$. As usual, the bandwidth has to go to zero but slower than the sample size grows to infinity. Also the bandwidth has to be chosen such that the asymptotic bias term vanishes faster than the variance. This allows to apply the Central Limit Theorem and makes the estimator asymptotically unbiased. In particular, condition (ii) requires undersmoothing in the sense that the bandwidth has to be below the mean squared error (MSE) optimal rate. As discussed for example in \textcite{Pagan_Ullah_1999}, choosing a higher order kernel mitigates the problem.\\
Conditions (iii) and (iv) state that $h$ has to be chosen such that first stage convergence rates vanish fast enough. In particular, by condition (iv) the joint convergence rates from propensity score and outcome projection estimation have to vanish faster than $\sqrt{N}$ scaled with the kernel bandwidth. Since $h\rightarrow 0$, this is a more restrictive assumption compared to the rate conditions usually required for average effects estimation (see for example \cite{Chernozhukov_Chetverikov_Demirer_Duflo_Hansen_Newey_2017}). In contrast to average effects, one is only interested in estimating the effect at a prespecified point $Z=z$. Thus, since sample observations enter the estimator in a weighted form, the prediction precision needed is for the lower effective sample size around $Z=z$. Hence, the first stage prediction guarantees need to adapt to this smaller sample conditions and therefore achieve a faster joint rate of convergence in terms of the sample size $N$. Condition (iii) additionally prevents the worst rate from becoming arbitrarily slow especially when $\lambda_Z$ is larger than one. Still, our estimator has a `rate' double robustness feature in the sense that joint rates can vanish relatively slowly but all single first stage rates are restricted from converging very slowly. $L_2$ convergence rates of many supervised machine learning methods satisfy these properties under sparsity conditions. For example \textcite{Belloni_Chernozhukov_2013} show that the predictive error of the Lasso is of order $O\left(\sqrt{\frac{s\log\max(\lambda_X,N)}{N}}\right)$ where $s$ the unknown number of true coefficients in the oracle model. Suppose that $s$ and $\lambda_X$ are equal in the outcome and the propensity score models then we require $\frac{s^2\log^2\max(\lambda_X,N)}{Nh^{\lambda_Z}}\rightarrow 0$. Hence the dimension of the confounding variables $\lambda_X$ can grow with the effective sample size $Nh^{\lambda_Z}$. Similar rates can be shown for $L_2$ boosting (\cite{Luo_Spindler_2016}) and nonlinear models like Random Forests (\cite{Wager_Walther_2015}) or forms of Deep Neural Nets (\cite{Farrell_Liang_Misra_2018}).\footnote{For the concrete dependence of sparsity conditions on the parameters of the predictors see the references mentioned.}\\
A natural question is then if a bandwidth exists that satisfies the rate conditions in Assumption \ref{ass:cate}. Indeed, one can show (for more details see Appendix \ref{app:range}) that the theoretical range of possible bandwidth choices can be described by
\begin{align*}
\frac{1}{\lambda_Z+2r}<\delta_h<\frac{2(\delta_{\epsilon_p}+\delta_{\epsilon_{m_d}})-1}{\lambda_Z}
\end{align*}
and we achieve a condition for the order of the kernel
\begin{align*}
r>\lambda_Z\frac{1-(\delta_{\epsilon_p}+\delta_{\epsilon_{m_d}})}{2(\delta_{\epsilon_p}+\delta_{\epsilon_{m_d}})-1}.
\end{align*}
For example if we restrict ourselves on second order kernel functions then for $\lambda_Z=1$ we require $\delta_{\epsilon_p}+\delta_{\epsilon_{m_d}}=\frac{3}{5}$. Similarly, for $\lambda_Z=2$ and $\lambda_Z=3$, $\delta_{\epsilon_p}+\delta_{\epsilon_{m_d}}=\frac{2}{3}$ and $\delta_{\epsilon_p}+\delta_{\epsilon_{m_d}}=\frac{5}{7}$ are required respectively. Thus, for a growing dimension $\lambda_Z$ the joint rate condition for the first stage nuisance parameters approaches the parametric rate.\\
The discussion indicates that given one has chosen an appropriate order of the kernel function, the researcher can choose the bandwidth somehow below but not too much below the MSE optimal rate. One could therefore simply use a certain fraction (e.g. 0.9) of the cross-validation bandwidth choice. Hence, from a practical perspective our bandwidth choice problem is equivalent to nonparametric regression with undersmoothing.\\
Given these assumptions we can then derive the first main theoretical result.\\
\begin{theorem}\label{thm:cate}
Under Assumptions \ref{ass:cia}-\ref{ass:cate} our proposed estimation procedure for GATE obeys
\begin{align*}
\sqrt{Nh^{\lambda_Z}}\left(\hat{\tau}-\tau\right)=\frac{1}{\sqrt{Nh^{\lambda_Z}}}\sum_{i=1}^N\frac{K\left(\frac{z_i-z}{h}\right)}{\frac{1}{Nh^{\lambda_Z}}\sum_{i=1}^NK\left(\frac{z_i-z}{h}\right)}\left(\psi(W_i,p,m_0,m_1)-\tau\right)+o(1)
\end{align*}
and
\begin{align*}
\sqrt{Nh^{\lambda_Z}}\left(\hat{\tau}-\tau\right)\rightarrow_d N(0,\sigma_{GATE}^2)
\end{align*}
with $\sigma_{\text{GATE}}^2=\frac{\int K(u)^2du\times\mathbb{E}\left[\left(\psi(W_i,p,m_0,m_1)-\tau\right)^2|Z=z\right]}{f(z)}$.\\
\end{theorem}
Theorem \ref{thm:cate} shows that under the assumptions discussed above the speed of convergence is determined only by the nonparametric regression step. In particular, it does not depend on the first stage estimation steps. An equivalent result can also be achieved by using IPW with nonparametric first stages (see \cite{Abrevaya_Hsu_Lieli_2015}). However, this requires an additional bandwidth choice for the first stage propensity score regression and is limited to the case when also $\lambda_X$ is very small. The dimension of $X$ can be increased under functional form assumptions for the first stage. However, as shown by the authors the price to pay is an increase in the asymptotic variance. This is not the case for the estimator proposed in this paper. Also Theorem \ref{thm:cate} is valid under generally weaker conditions compared to the results in \textcite{Lee_Okui_Whang_2016} for parametric first stages. Heuristically\footnote{In contrast, to \textcite{Lee_Okui_Whang_2016} we use local constant instead of local linear regression and introduce cross-fitting for nuisance parameter estimation. This should, however, not be a concern for the intuitive argument made.}, if the first stage estimators converge at $\sqrt{N}$ then our conditions on the bandwidth are satisfied and our asymptotic results continue to apply. To this extent, our results comprise the result of \textcite{Lee_Okui_Whang_2016} as a special case.
\subsection{Joint estimation of GATE and ATE}
\subsubsection{Proposed estimator}
Given the considerations so far, it appears `naturally' to estimate ATE in three steps as an average of GATEs in the sample. The details of the proposed estimator are described in Procedure 2.
\begin{figure}[h]
\centering
\onehalfspacing
\fbox{\begin{minipage}{0.95\textwidth}
\textbf{Procedure 2.} ATE estimation\\~\\
\begin{enumerate}
\item Follow steps 1-3 of Procedure 1.
\item Predict GATE at every observation in the sample as $\hat{\tau}(z_j)$.
\item Estimate ATE as
\begin{align*}
\hat{\theta}=\frac{1}{N}\sum_{j=1}^N\hat{\tau}(z_j).
\end{align*}
\end{enumerate}
\end{minipage}}
\end{figure}
As suggested in \textcite{Chernozhukov_Chetverikov_Demirer_Duflo_Hansen_Newey_2017} one could also directly estimate ATE as the average of the vector $\psi(W,p,m_0,m_1)$ using the first stage nuisance parameter predictions. However, as a by-product of GATE estimation, using an additional kernel smoothing step may lead to an ATE estimator with better finite sample properties. In particular, the propensity score weights do not enter the last step of our estimator directly and our hope is that small misspecification errors of propensity scores close to zero or one are therefore smoothed out. The sensitivity of estimators incorporating inverse propensity score weights directly is the subject of many Monte Carlo experiments (e.g. \cite{Huber_Lechner_Wunsch_2013}, and \cite{Froelich_2004b}). We notice that similar reasoning is also behind three-step estimators that apply nonparametric regression on an estimated propensity score often used in practice.
\subsubsection{Asymptotic results}
To obtain our theoretical results we have to modify Assumption \ref{ass:cate} slightly.\\
\begin{assumptionprime}[Coupled convergence (ATE)]\label{ass:ate}~
The bandwidth $h$ and the sample size $N$ jointly converge such that
\begin{itemize}
\item[(i)] $h=o(1)$, $Nh^{\lambda_Z}\rightarrow\infty$ as $N\rightarrow\infty$,
\item[(ii)] $N^{\frac{1}{2}}h^{\frac{1}{2}\lambda_Z}h^r=o(1)$ and
\item[(iii)] $Nh^{4r}=o(1)$ and $Nh^{2\lambda_Z}\rightarrow\infty$.
\end{itemize}
Further, $N$ and $h$ satisfy the joint convergence conditions with the nuisance parameter convergence rates
\begin{itemize}
\item[(iv)] $h^{-\lambda_Z}\epsilon_{\max}=o(1)$ and
\item[(v)] $N^{\frac{1}{2}}h^{-\lambda_Z}\epsilon_{m_0}\epsilon_p+N^{\frac{1}{2}}h^{-\lambda_Z}\epsilon_{m_1}\epsilon_p=o(1)$.\\
\end{itemize}
\end{assumptionprime}
We notice that averaging over the estimated projection $\mathbb{E}\left[\psi(W,p,m_0,m_1)|X\right]$ is a partial mean problem in the sense of \textcite{Newey_1994b, Newey_1994}. While parts (i) and (ii) of Assumption \ref{ass:cate} remain unchanged, the additional condition (iii) is necessary in order to guarantee that the MSE of the kernel regression estimator scaled with $N^{\frac{1}{4}}$ converges to zero. In this way we guarantee the applicability of \posscite{Newey_1994} framework. We could have also assumed uniform convergence rates for the kernel regression step. However, this would involve a unnecessarily strong condition (for a discussion see \cite[pp. 1364-1368]{Newey_1994} and also \cite[p. 2205]{Newey_McFadden_1994}).\\
Since we want ATE to converge with a rate of $\sqrt{N}$, the requirements on the first stage convergence rates in condition (v) are more restrictive than those in the respective condition of Assumption \ref{ass:cate}. The range of theoretically feasible bandwidth choices reduces to
\begin{align*}
\max\left(\frac{1}{4r},\frac{1}{\lambda_Z+2r}\right)<\delta_h<\frac{(\delta_{\epsilon_p}+\delta_{\epsilon_{m_d}})-\frac{1}{2}}{\lambda_Z}.
\end{align*}
Assuming $\frac{1}{4r}<\frac{1}{\lambda_Z+2r}$ we get a modified condition for the order of the kernel function
\begin{align*}
r>\lambda_Z\frac{\frac{3}{2}-(\delta_{\epsilon_p}+\delta_{\epsilon_{m_d}})}{2(\delta_{\epsilon_p}+\delta_{\epsilon_{m_d}})-1}.
\end{align*}
In general this result indicates that one relies on a higher-order kernel function whenever $\lambda_Z>1$ when GATE and ATE are estimated jointly.\\
Under the stronger Assumption \ref{ass:ate} we can then derive the following efficiency result.\\
\begin{theorem}\label{thm:ate}
Under Assumptions \ref{ass:cia}-\ref{ass:superror} and \ref{ass:ate} and the regularity conditions on the nonparametric second step as in \textcite[pp. 1364-1368]{Newey_1994} our proposed estimation procedure for ATE has the influence function
\begin{align*}
\frac{D(Y-m_1(X))}{p(X)}-\frac{(1-D)(Y-m_0(X))}{1-p(X)}+m_1(X)-m_0(X)-\theta
\end{align*}
and therefore obeys
\begin{align*}
\sqrt{N}(\hat{\theta}-\theta)\rightarrow_d N(0,\sigma_{ATE}^2)
\end{align*}
where $\sigma_{ATE}^2$ is the semiparametric efficiency bound of \textcite{Hahn_1998}.\\
\end{theorem}
Conceptually, Theorem \ref{thm:ate} underpins our intuition from semiparametric theory outlined in Section \ref{subsec:semiparametric}. The result shows that indeed every estimator that involves a nonparametric projection of the AIPW modified outcome on any low-dimensional subset of $X$ is consistent, asymptotically normal and achieves the semiparametric efficiency bound. This asymptotic result has also been shown for other estimators already discussed in Section \ref{sec:introduction}. In contrast to \posscite{Hirano_Imbens_Ridder_2003} estimator, \posscite{Hahn_1998} estimator and matching on the propensity score (\cite{Hahn_Ridder_2013}), we do not rely on nonparametrically estimated first stages. Due to the fact that $\lambda_X\rightarrow\infty$ these estimators are of no practical use in our setting. Further, unlike AIPW with machine learning nuisance parameter estimation (\cite{Chernozhukov_Chetverikov_Demirer_Duflo_Hansen_Newey_2017}), our estimator involves an additional step. Therefore the inverse propensity score does not directly enter our estimator but is smoothed through the additional nonparametric step. Asymptotically, this does not make any difference as the result in Theorem \ref{thm:ate} shows. However, in finite sample this could be a major advantage over the usual AIPW estimator.
\section{Illustrative example}\label{sec:example}
We investigate the applicability of our methods using \posscite{Cattaneo_2010} dataset on the effect of cigarette smoking on birthweight available from the Stata website.\footnote{The original dataset can be retrieved from \href{http://www.stata-press.com/data/r13/cattaneo2.dta}{here}.} The dataset contains the outcome variable birthweight in grams ($Y$), whether the mother smoked during pregnancy ($D=1$) and several covariates on the mother's health and socio-economic background ($X$). A detailed description of all covariates in the dataset can be found in Appendix \ref{app:data}. Applied studies with different estimation approaches unambiguously find negative average effects (see \cite{Abrevaya_2006}, \cite{daVeiga_Wilder_2008}, \cite{Walker_Tekin_Wallace_2009}). Conditional average treatment effects were investigated by \textcite{Abrevaya_Hsu_Lieli_2015} and \textcite{Lee_Okui_Whang_2016} who find that mother's age is associated with increasingly negative effects of smoking. We replicate their results and compare their estimators with ours. Clearly, this type of analysis is limited in its scope since the true DGP remains unknown. However, the dataset has the particular advantage that some strong hypothesis about the estimation results are plausible. (i) The effect of smoking on birthweight should be either negative or zero. (ii) The effect should be increasingly negative with mother's age.\\
As a second example we consider how the effect changes with the number of prenatal care visits. On the one hand a very low number of care visits could indicate the mother's insufficient access to medical infrastructure and therefore could be associated with particularly negative effects. On the other hand a very high number of care visits could indicate a poor health situation. Hence, it is a priori unclear how the treatment effect and health care visits are exactly related.
\subsection{Empirical results}
\begin{figure}[p]
\centering
\caption{AIPW GATE estimator with ensemble first stages}
\begin{subfigure}{0.49\textwidth}
\includegraphics[width=\textwidth]{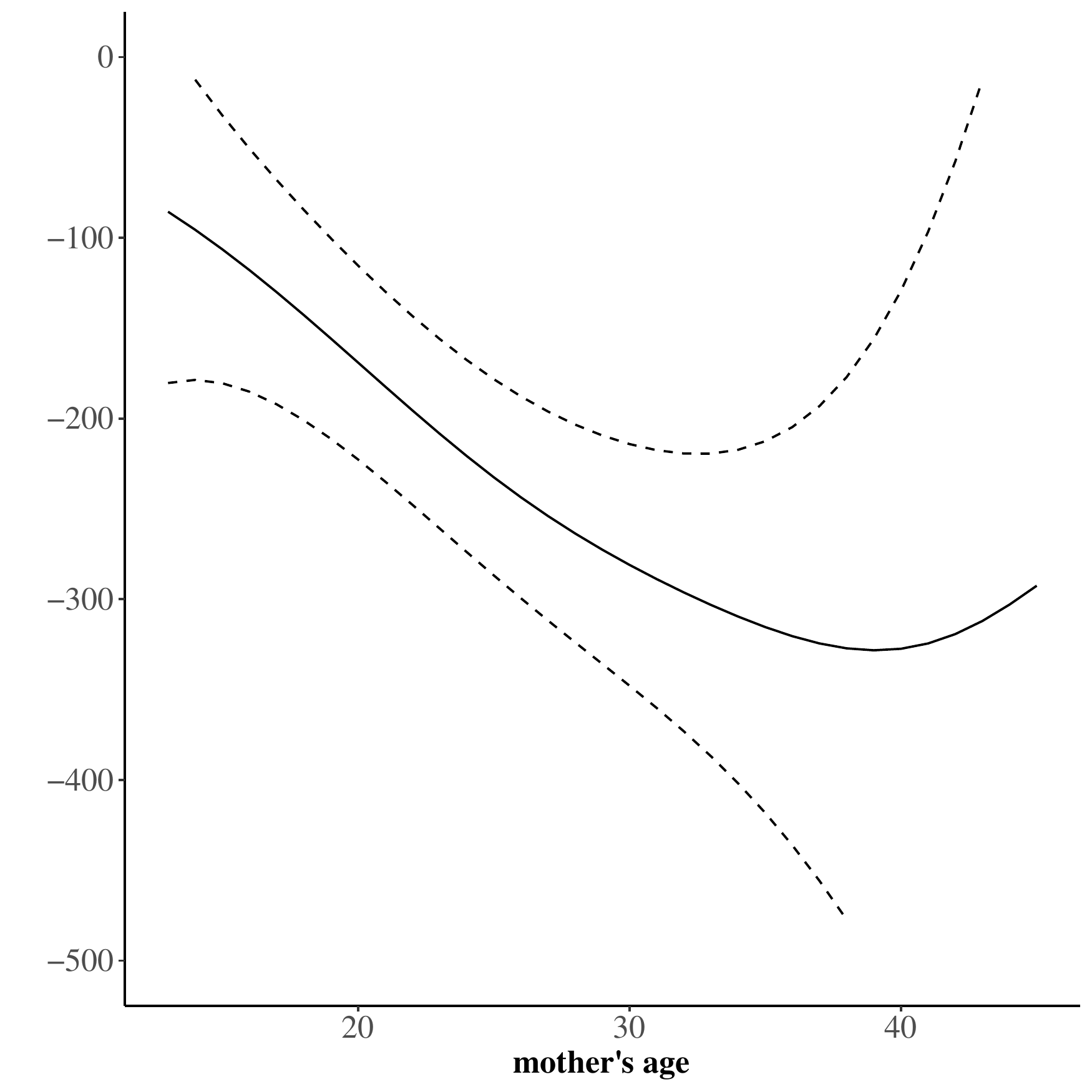}
\end{subfigure}
\begin{subfigure}{0.49\textwidth}
\includegraphics[width=\textwidth]{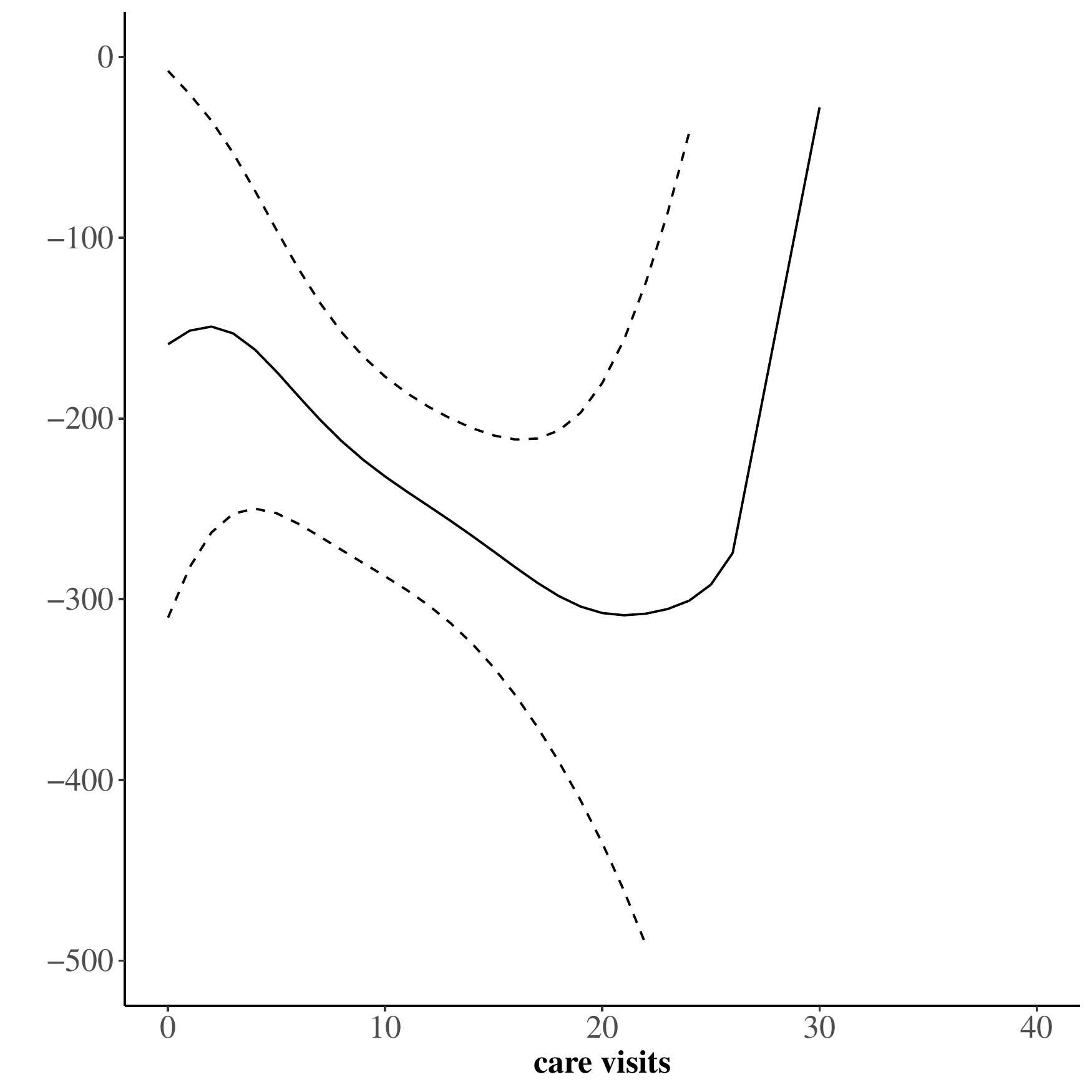}
\end{subfigure}
\begin{subfigure}{0.8\textwidth}
\includegraphics[width=\textwidth]{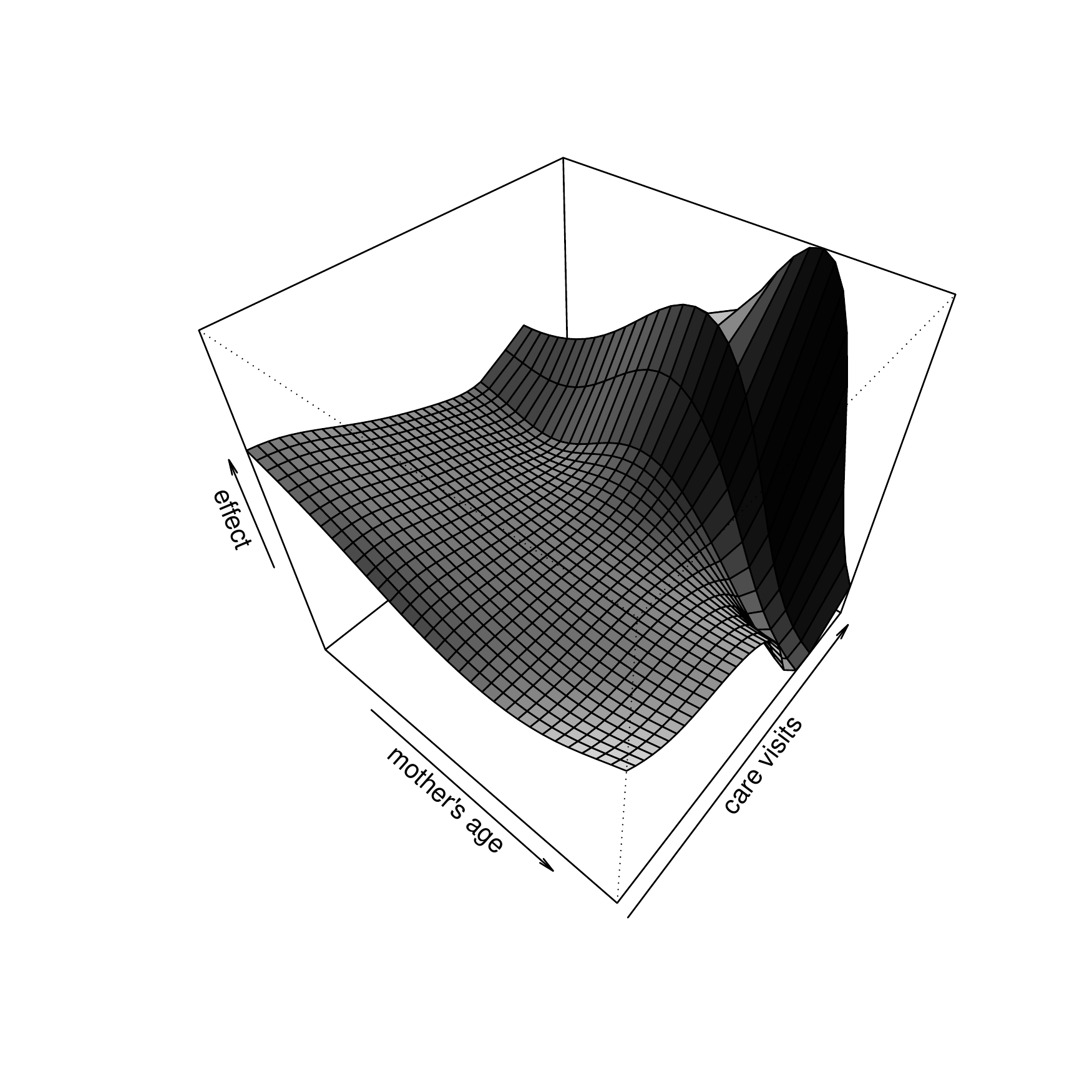}
\end{subfigure}
\label{fig:aipwcate}
\floatfoot{Results were obtained as described in Procedure 1 with a second-order Gaussian kernel function and a $0.9\times$LOOCV bandwidth choice. Nuisance parameters were estimated using an ensemble learner comprising Lasso, Elastic Net, Ridge and Random Forest. For Lasso, Ridge and Elastic Net the penalty term was chosen such that the cross-validation criterion was minimized. The ensemble weights were chosen by minimizing out-of-sample MSE. Asymptotic confidence bands are at the $95\%$ level.}
\end{figure}
\begin{figure}[h]
\centering
\caption{Sensitivity to bandwidth choice (age)}
\begin{subfigure}{0.3\textwidth}
\caption{$0.5$ $\times$ CV choice}
\includegraphics[width=\textwidth]{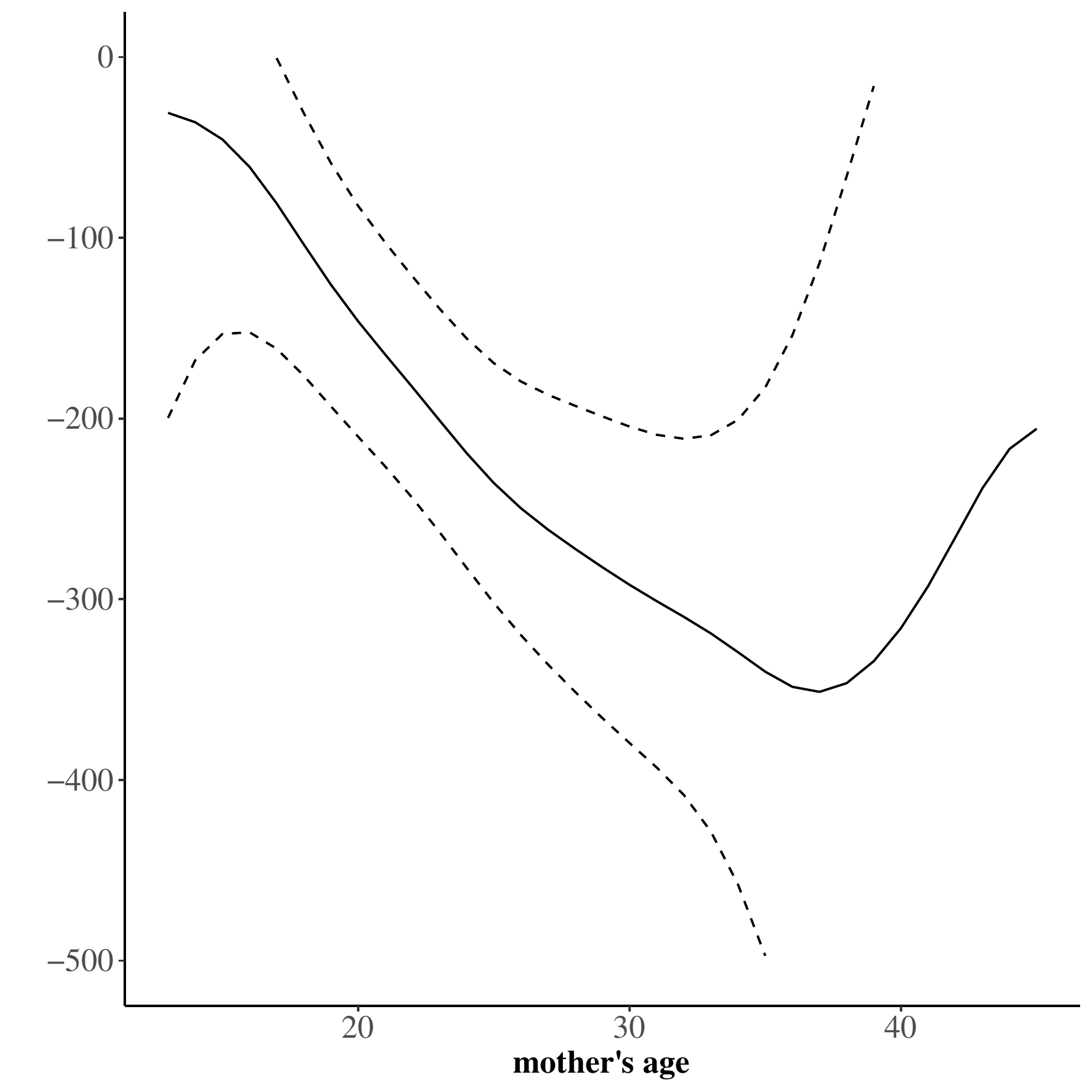}
\end{subfigure}
\begin{subfigure}{0.3\textwidth}
\caption{$0.7$ $\times$ CV choice}
\includegraphics[width=\textwidth]{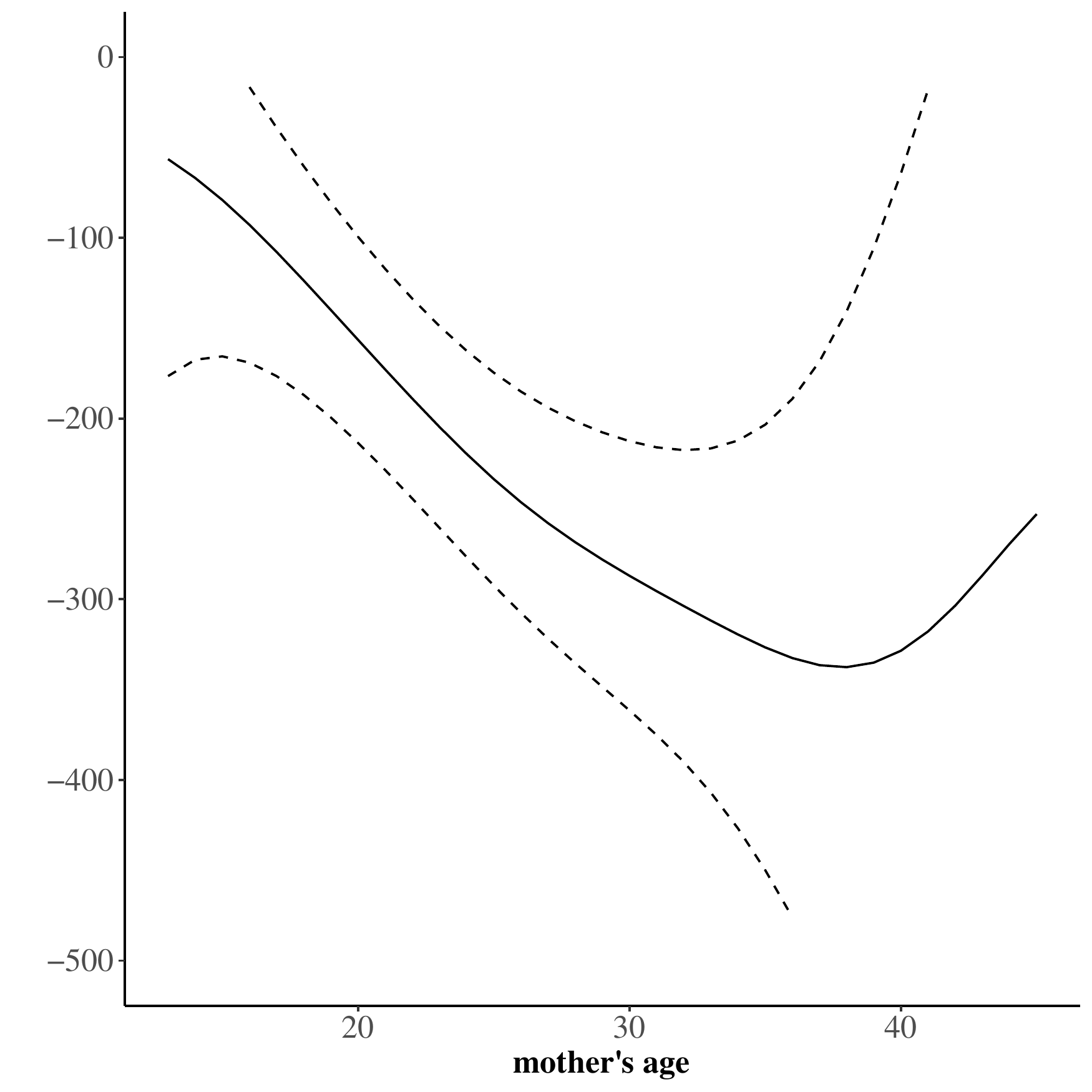}
\end{subfigure}
\begin{subfigure}{0.3\textwidth}
\caption{$0.8$ $\times$ CV choice}
\includegraphics[width=\textwidth]{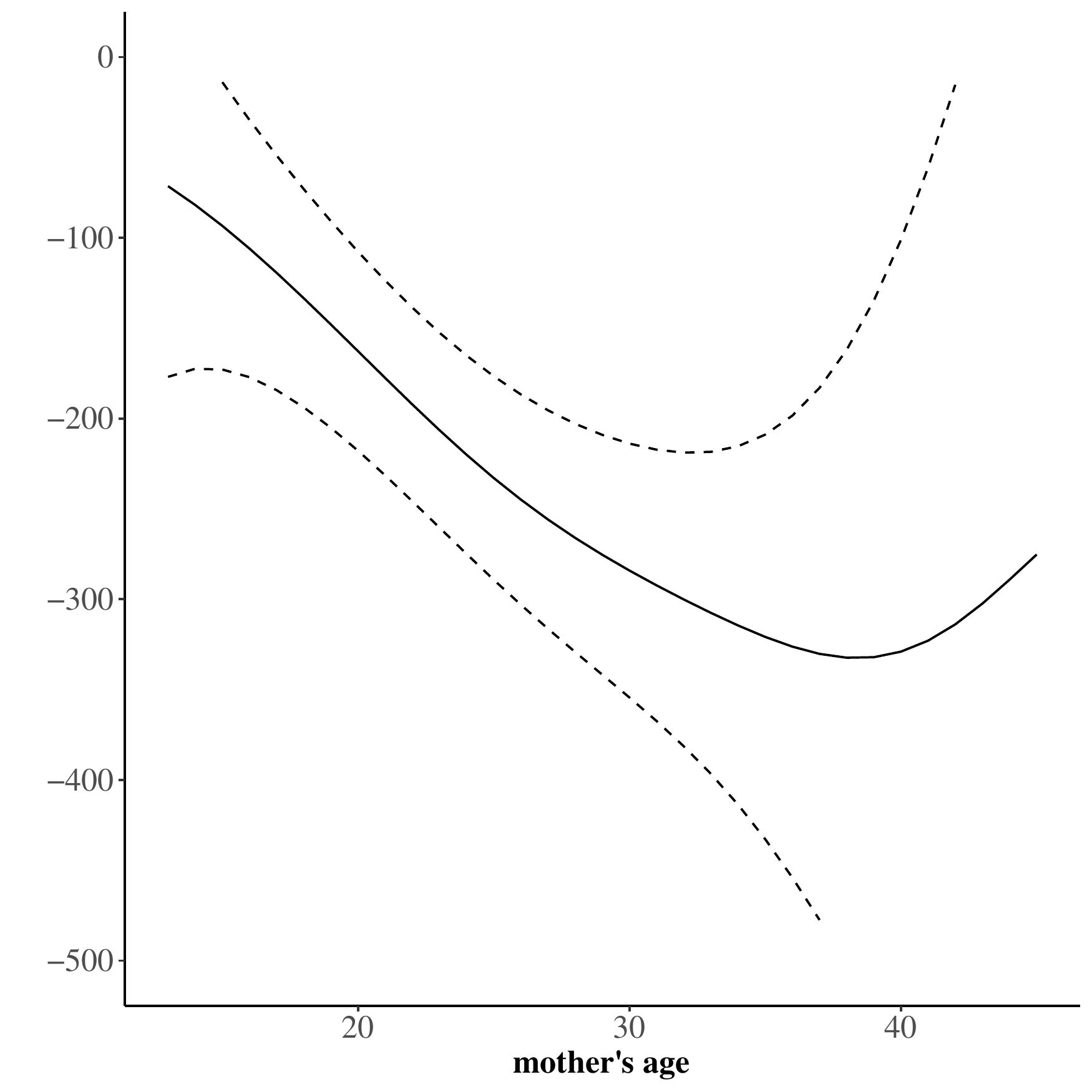}
\end{subfigure}
\begin{subfigure}{0.3\textwidth}
\caption{$0.9$ $\times$ CV choice}
\includegraphics[width=\textwidth]{figures/aipwvec_en_age.pdf}
\end{subfigure}
\begin{subfigure}{0.3\textwidth}
\caption{$1.0$ $\times$ CV choice}
\includegraphics[width=\textwidth]{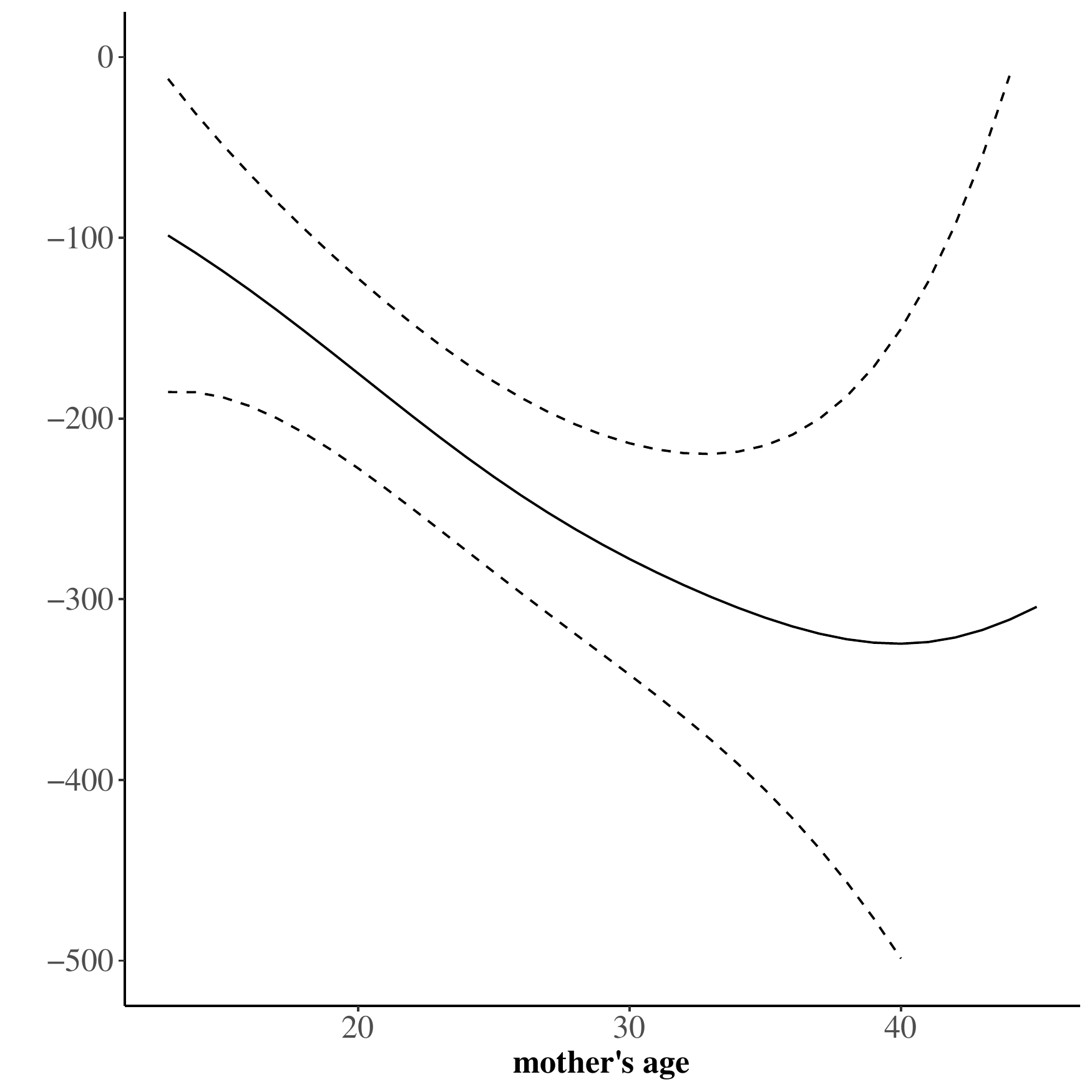}
\end{subfigure}
\begin{subfigure}{0.3\textwidth}
\caption{$1.5$ $\times$ CV choice}
\includegraphics[width=\textwidth]{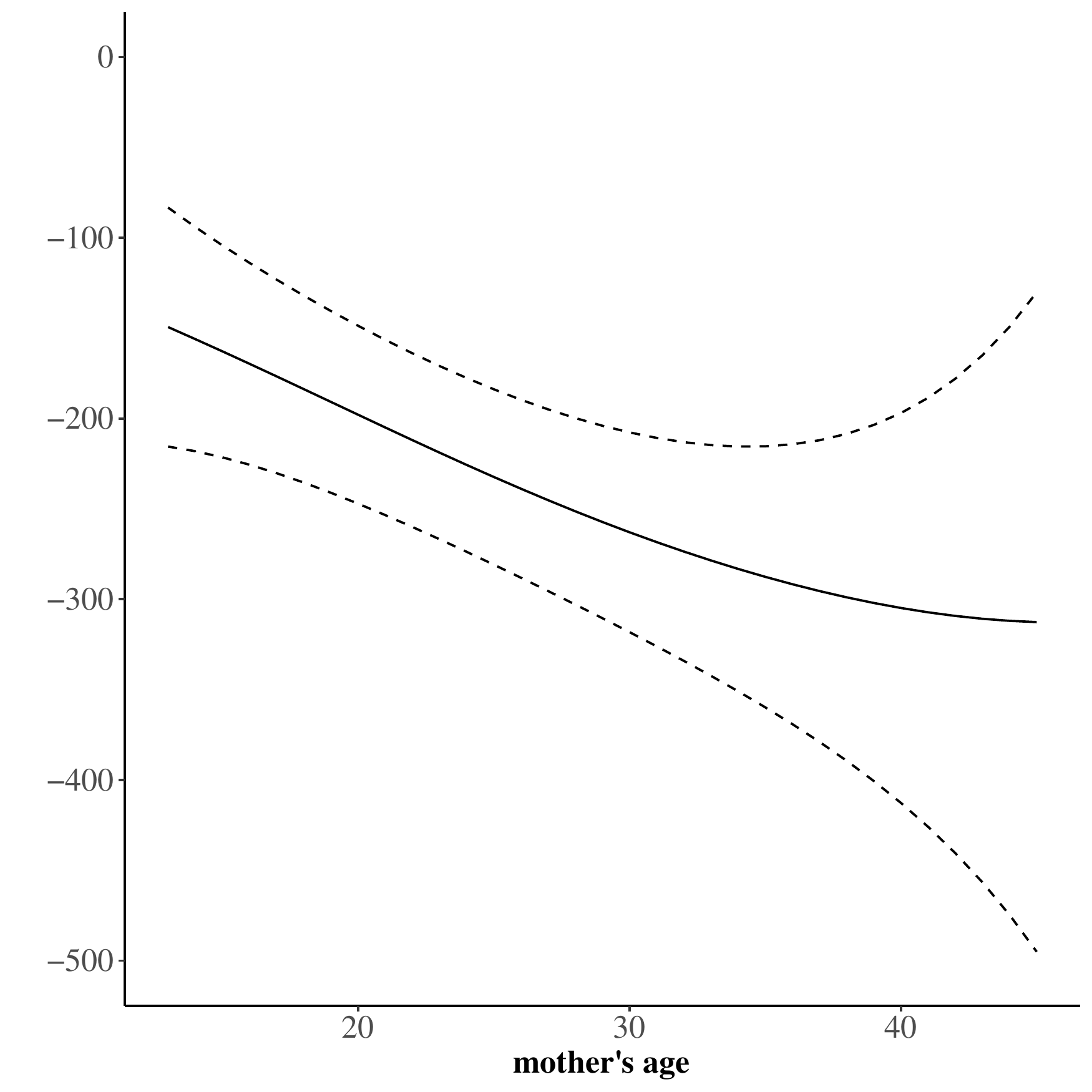}
\end{subfigure}
\label{fig:aipwbw}
\floatfoot{Results were obtained as described in Procedure 1 with a second-order Gaussian kernel function and different multiples of the LOOCV bandwidth choice. Nuisance parameters were estimated using an ensemble learner comprising Lasso, Elastic Net, Ridge and Random Forest. For Lasso, Ridge and Elastic Net the penalty term was chosen such that the cross-validation criterion was minimized. The ensemble weights were chosen by minimizing out-of-sample MSE. Asymptotic confidence bands are at the $95\%$ level.}
\end{figure}
Figure \ref{fig:aipwcate} depicts the main results of our empirical analysis. We estimate the GATEs as described in Procedure 1 using an ensemble learner comprising Lasso, Ridge, Elastic Net and a Random Forest. The weights of the ensemble are obtained by cross-validating the out-of-sample MSE of the procedure. $X$ in our specification is an extended variable set (`alldata') and is exactly documented in Appendix \ref{app:data}. For example in contrast to \textcite{Lee_Okui_Whang_2016} we also include the available characteristics for the father of the child, since they could be a good predictor for the smoking behaviour of the mother. The covariates enter our model very flexibly. For the penalized regression predictors we allow for polynomials up to order four and all two way interactions. The Random Forest has the particular advantage of being an ensemble of trees itself and is therefore very flexible by construction. The results are generally in line with the hypothesis made. In particular, the effect of smoking is unambiguously negative over the whole support of mother's age and prenatal care visits. As expected the effect increases with age. Interestingly, a higher number of prenatal care visits seems to be associated with higher negative effects.\\
We estimate all our results with second-order Gaussian kernel functions. The same analysis using higher order Gaussian kernel functions as proposed by \textcite{Li_Racine_2007} yields similar results (see Appendix \ref{app:data}). In practice the biggest challenge is to determine the bandwidth for the nonparametric regression. To achieve undersmoothing, we multiply the bandwidth obtained by leave-one-out cross-validation with 0.9. Since this choice is arbitrary, the stability of our results towards this choice is a particular concern. Figure \ref{fig:aipwbw} shows that our estimator is relatively robust regarding this choice. A major change in the shape of the function only appears for massive oversmoothing. An equivalent analysis for prenatal care visits yielding the same conclusion is relegated to Appendix \ref{app:data}.
\begin{table}[h]
\centering
\caption{Smoothed ATE estimators}
\label{tab:ateestsmooth}
\begin{threeparttable}
\begin{tabular}{ccc}
  \toprule
Smoothed AIPW (age) & Smoothed AIPW (care visits) & Smoothed AIPW (age, care visits)\\ 
  \midrule
-238.937 & -235.672 & -236.904\\
(27.257) & (27.257) & (27.257)\\
\bottomrule \bottomrule
\end{tabular}
\begin{tablenotes}
\small 
\item Results for smoothed AIPW ATE estimation as in Procedure 2 using $Z=\text{age}$, $Z=\text{care visits}$ and $Z=(\text{age},\text{care visits})$. Results were obtained with a second-order Gaussian kernel function and a $0.9\times$LOOCV bandwidth choice. Nuisance parameters were estimated using an ensemble learner comprising Lasso, Elastic Net, Ridge and Random Forest. For Lasso, Ridge and Elastic Net the penalty term was chosen such that the cross-validation criterion was minimized. The ensemble weights were chosen by minimizing out-of-sample MSE. Asymptotic standard errors are in parenthesis.
\end{tablenotes}
\end{threeparttable}
\end{table}
\\
Finally, Table \ref{tab:ateestsmooth} shows the results for ATE estimation as described in Procedure 2. In line with the previous literature mentioned above, the average effect of smoking is estimated to be negative. Crucially, the estimated effect turns out to be very robust regarding the choice of the smoothing variable.
\subsection{Comparison with other estimators}
\begin{figure}[p]
\centering
\caption{Other GATE estimators}
\begin{subfigure}{0.49\textwidth}
\caption{AIPW linear first stages (age)}
\includegraphics[width=\textwidth]{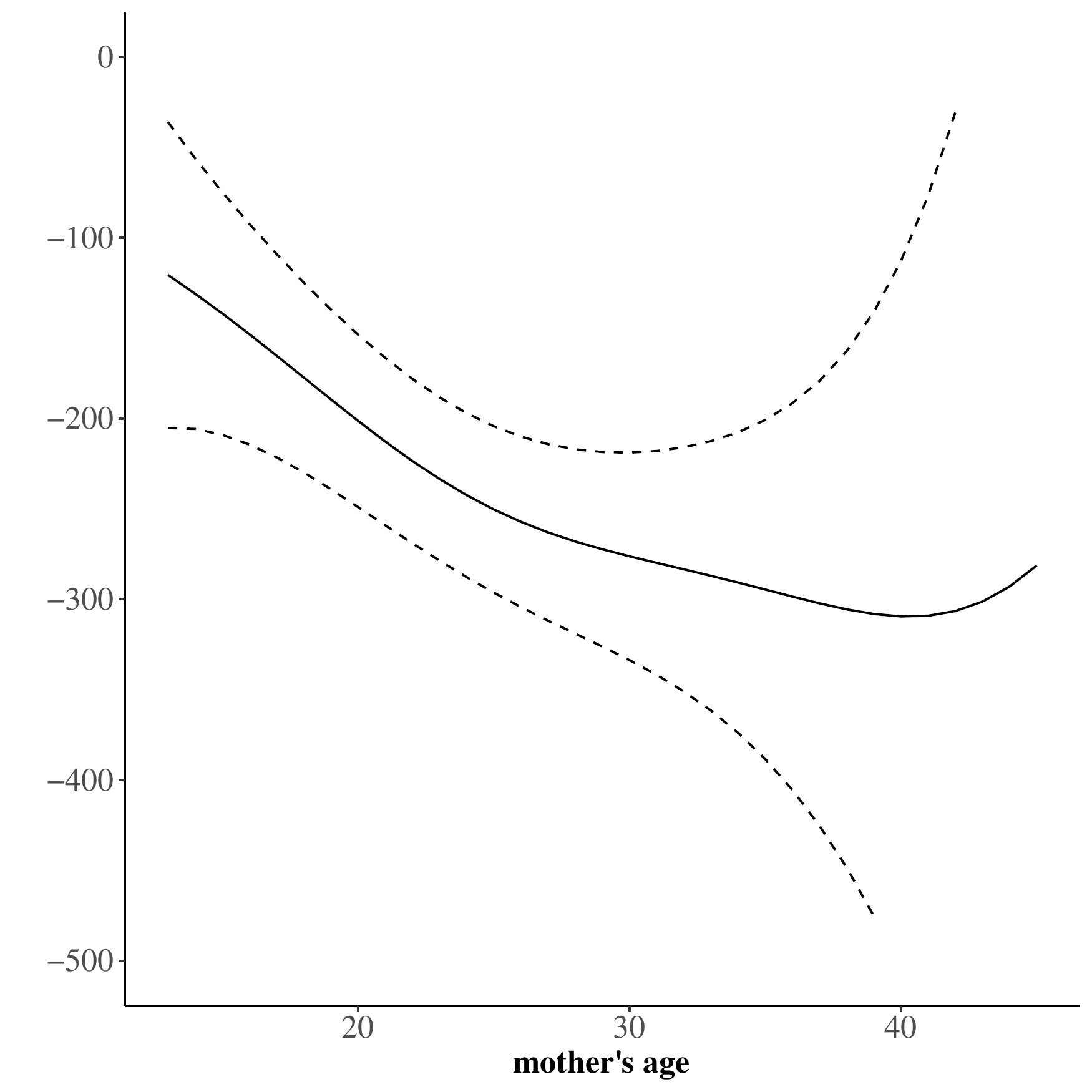}
\end{subfigure}
\begin{subfigure}{0.49\textwidth}
\caption{AIPW linear first stages (care visits)}
\includegraphics[width=\textwidth]{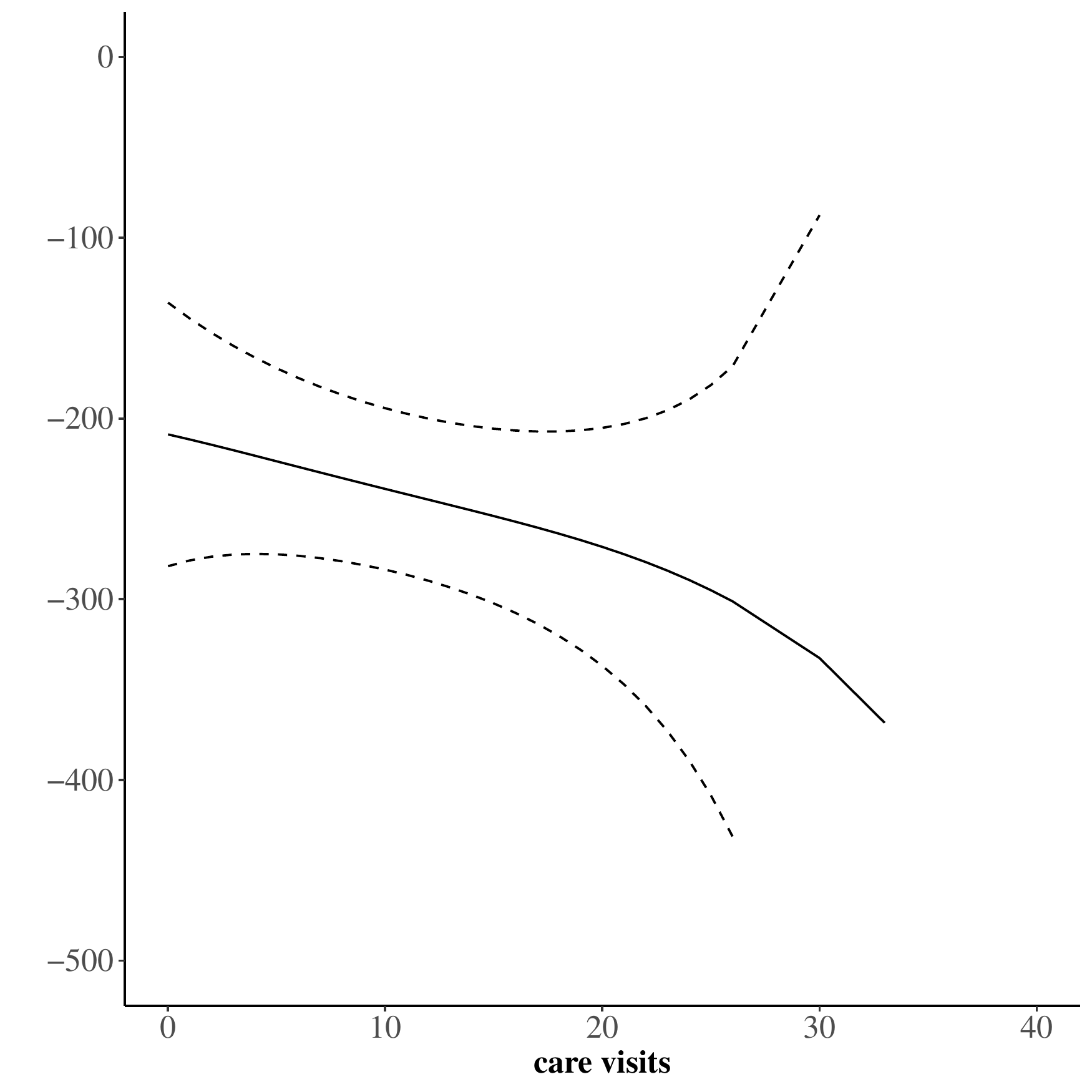}
\end{subfigure}
\begin{subfigure}{0.49\textwidth}
\caption{IPW linear first stages (age)}
\includegraphics[width=\textwidth]{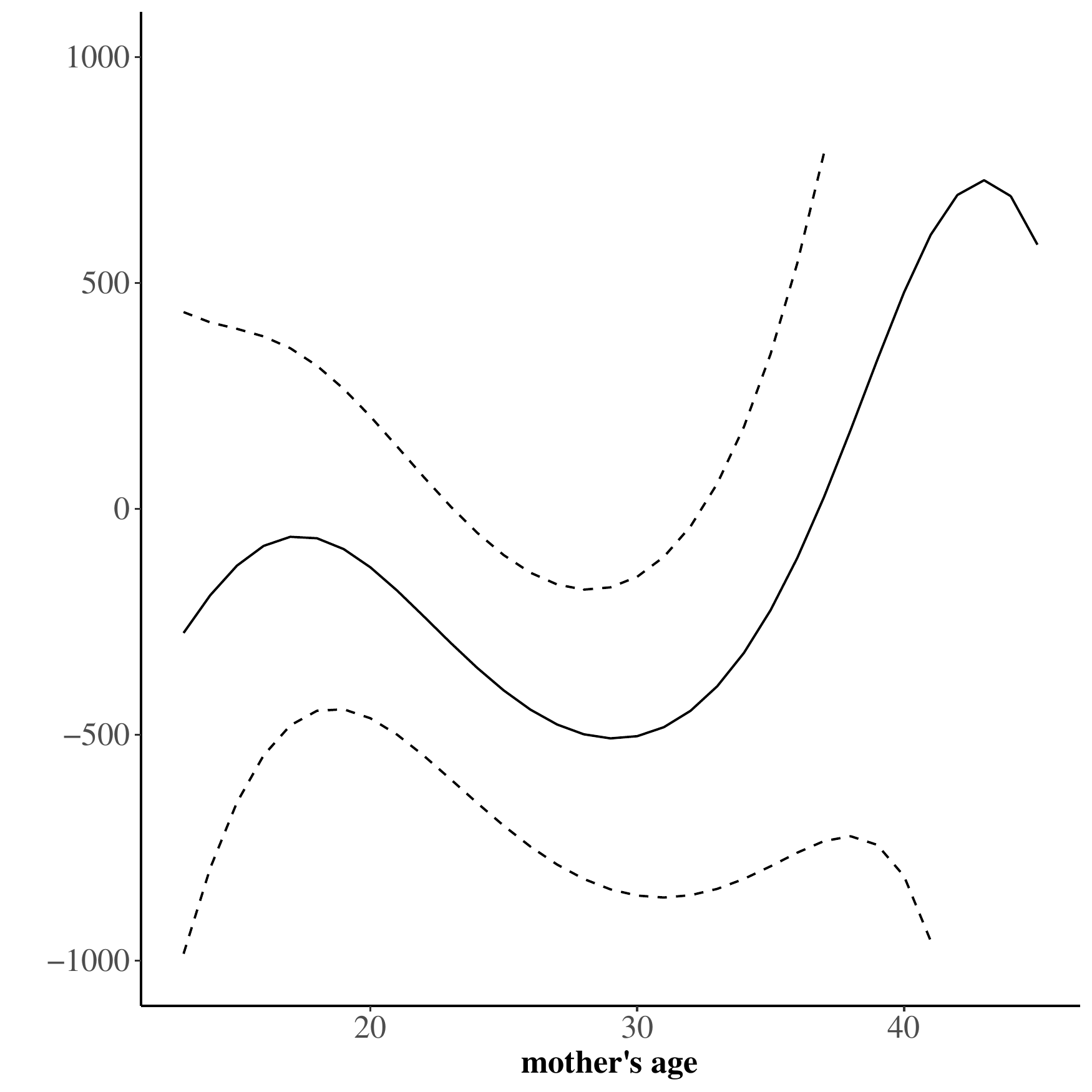}
\end{subfigure}
\begin{subfigure}{0.49\textwidth}
\caption{IPW linear first stages (care visits)}
\includegraphics[width=\textwidth]{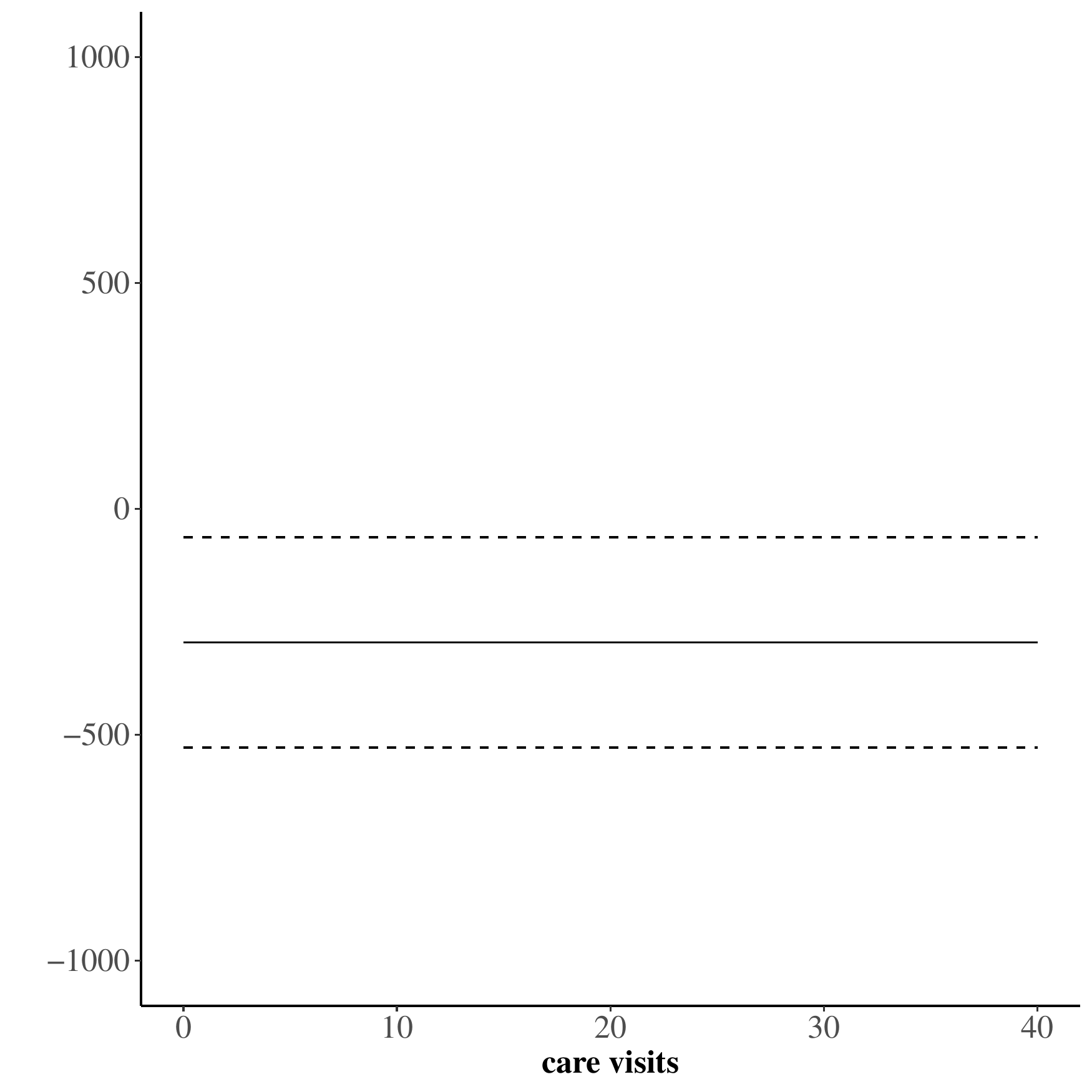}
\end{subfigure}
\label{fig:othercate}
\floatfoot{Results were obtained following the procedures in \textcite{Abrevaya_Hsu_Lieli_2015} and \textcite{Lee_Okui_Whang_2016} with a second-order Gaussian kernel function and a $0.9\times$LOOCV bandwidth choice. Nuisance parameters were estimated using Logit for the propensity score and OLS for the outcome projections. Asymptotic confidence bands are at the $95\%$ level.}
\end{figure}
A `fair' comparison with other estimators is hardly feasible because our approach does not require specific functional form assumptions.\footnote{We do not consider nonparametric propensity score estimation as suggested in \textcite{Abrevaya_Hsu_Lieli_2015} because it most likely does not allow to include all potential confounders in order to make Assumption \ref{ass:cia} credible.} In other words, the related estimators of \textcite{Abrevaya_Hsu_Lieli_2015} and \textcite{Lee_Okui_Whang_2016} suppose that they know the true propensity score or outcome projection specifications. Since we cannot compare our estimator against every possible parametric specification, we use the specification selected by \textcite{Lee_Okui_Whang_2016} as a benchmark. Figure \ref{fig:othercate} depicts GATE estimation results using the benchmark models. Strikingly, the IPW based estimator gives implausible results. For mother's age positive effects of smoking can almost nowhere be excluded. For care visits we do not obtain a GATE estimation result for our bandwidth choice. In fact, ATE is estimated indicating that the bandwidth is too large in order to obtain GATE estimates. In line with the theoretical result, standard errors are inflated compared to our estimation procedure. The AIPW based estimator with parametric models for the propensity score and the outcome projections gives plausible results for GATE with respect to mother's age. Slight differences arise when comparing the GATE curves regarding the number of prenatal care visits.
\begin{table}[h]
\centering
\caption{Averaged ATE estimators}
\label{tab:ateest}
\begin{threeparttable}
\begin{tabular}{ccc}
  \toprule
Averaged AIPW (ensemble) & Averaged AIPW (parametric) & Averaged IPW (parametric)\\ 
  \midrule
 -234.826 & -242.990 & -295.388 \\
(27.257) & (25.885) & (45.110)\\
\bottomrule \bottomrule
\end{tabular}
\begin{tablenotes}
\small 
\item Results for ATE estimation using Inverse Probability Weighting (IPW) and Augmented IPW (AIPW). For the ensemble learner nuisance parameters were estimated using an ensemble learner comprising Lasso, Elastic Net, Ridge and Random Forest. For Lasso, Ridge and Elastic Net the penalty term was chosen such that the cross-validation criterion was minimized. The ensemble weights were chosen by minimizing out-of-sample MSE. For the parametric specifications nuisance parameters were estimated using Logit for the propensity score and OLS for the outcome projections. Asymptotic standard errors are in parenthesis.
\end{tablenotes}
\end{threeparttable}
\end{table}
\\
Table \ref{tab:ateest} shows the results for ATE estimation. As expected the results of Procedure 2 in Table \ref{tab:ateestsmooth} are roughly in line with the standard AIPW based ATE estimator with ensemble first stages.\footnote{This might also be seen as a implicit test for the credibility of the stronger conditions required for the smoothed estimator compared to the averaged efficient score as in \textcite{Chernozhukov_Chetverikov_Demirer_Duflo_Hansen_Newey_2017}.} The relative bad performance of IPW based estimation for the GATEs is also reflected in the estimation of ATE. In particular, the standard error nearly doubles compared to AIPW based estimators and point estimates are reduced. Interestingly, for average effects there seems to be only little value-added for the flexible machine learning based estimators compared to the parametric specification.
\section{Conclusion}
In this study we propose new estimators for specific conditional and average causal effects when the dimension of the covariate space is high. In particular, by discriminating the different roles of covariates (adjusting for confounding vs. measuring causal heterogeneity of interest) in our approach, they can be included very flexibly -- not relying on any functional form assumptions. Rather, we show coupled convergence conditions for the different steps involved. The procedures suggested are based on semiparametric efficiency theory. In this sense, our proposed three-step estimator for ATE estimation is shown to reach the semiparametric efficiency bound. A widely used empirical example shows that our estimators are useful in practice. Compared to other estimators their desirable theoretical properties and increased flexibility could lead to divergent empirical results.\\
The specific structure of the GATE problem should be easily applicable to related settings. For example efficient score based estimation can also be used for instrumental variables problems (\cite{Chernozhukov_Chetverikov_Demirer_Duflo_Hansen_Newey_2017}), difference-in-differences estimation (\cite{Zimmert_2018}) and continuous treatment settings (\cite{Kennedy_Ma_McHugh_Small_2017}).\\
Some other interesting problems and refinements are beyond the scope of this study and have to be left for further research as well. For example, the nonparametric regression estimator could be refined to the extent that its bandwidth is chosen in a data-adaptive manner. As an alternative to classical nonparametric regression, one could also investigate using methods from the toolbox of supervised machine learning. This might help to get reliable estimators even for cases when the dimension of $Z$ is moderately higher than considered in this paper, while sacrificing only little flexibility.\\
Finally, it might be worth to investigate the finite sample properties of the proposed three-step estimators for ATE compared to averaging the efficient score vector directly. Here, we consider our ATE estimator as a by-product of the GATE procedure underpinning the theoretical motivation of our framework. While the smoothed three-step estimator is first order asymptotically equivalent to directly averaging the efficient score vector, it might posses better finite sample properties since it does not directly rely on propensity score weights. However, the finite sample performance may crucially rely on the bandwidth choice and the set of covariates in $Z$. We regard this as yet another interesting direction for further research.
\newpage
\printbibliography
\newpage
\appendix
\section{Proof of Theorems}
\subsection{Proof of Theorem \ref{thm:cate}}
We can write
\begin{align*}
\hat{\tau}-\tau=&\frac{\frac{1}{Nh^{\lambda_Z}}\sum_{i=1}^NK\left(\frac{z_i-z}{h}\right)\left(\psi(W_i,\hat{p},\hat{m}_0,\hat{m}_1)\right)-\tau}{\frac{1}{Nh^{\lambda_Z}}\sum_{i=1}^NK\left(\frac{z_i-z}{h}\right)}\\
=&\underbrace{\frac{\frac{1}{Nh^{\lambda_Z}}\sum_{i=1}^NK\left(\frac{z_i-z}{h}\right)\left(\psi(W_i,p,m_0,m_1)-\tau\right)}{\frac{1}{Nh^{\lambda_Z}}\sum_{i=1}^NK\left(\frac{z_i-z}{h}\right)}}_{i}\\
&+\underbrace{\frac{\frac{1}{Nh^{\lambda_Z}}\sum_{i=1}^NK\left(\frac{z_i-z}{h}\right)\left(\psi(W_i,\hat{p},\hat{m}_0,\hat{m}_1)-\psi(W_i,p,m_0,m_1)\right)}{\frac{1}{Nh^{\lambda_Z}}\sum_{i=1}^NK\left(\frac{z_i-z}{h}\right)}}_{ii}
\end{align*}
\subsubsection*{Influence function}
Denote $\bar{\psi}_i=\psi(W_i,\hat{p},\hat{m}_0,\hat{m}_1)-\psi(W_i,p,m_0,m_1)$. Then the second term can be further expanded as
\begin{align*}
ii=\underbrace{\frac{\frac{1}{Nh^{\lambda_Z}}\sum_{i=1}^N\mathbb{E}\left[K\left(\frac{Z-z}{h}\right)\bar{\psi}\right]}{\frac{1}{Nh^{\lambda_Z}}\sum_{i=1}^NK\left(\frac{z_i-z}{h}\right)}}_{iia}+\underbrace{\frac{\frac{1}{Nh^{\lambda_Z}}\sum_{i=1}^N\left(K\left(\frac{z_i-z}{h}\right)\bar{\psi}_i-\mathbb{E}\left[K\left(\frac{Z-z}{h}\right)\bar{\psi}\right]\right)}{\frac{1}{Nh^{\lambda_Z}}\sum_{i=1}^NK\left(\frac{z_i-z}{h}\right)}}_{iib}
\end{align*}
and therefore
\begin{align*}
\left\lvert \sqrt{Nh^{\lambda_Z}}ii\right\rvert\leq \left\lvert \sqrt{Nh^{\lambda_Z}}iia\right\rvert + \left\lvert \sqrt{Nh^{\lambda_Z}}iib\right\rvert.
\end{align*}
\textit{Bounding iia}\\
We first of all notice that
\begin{align*}
\left\lvert \sqrt{Nh^{\lambda_Z}}iia\right\rvert \leq \left\lvert\hat{f}(z)^{-1}\right\rvert\times\left\lvert\frac{\sqrt{N}}{\sqrt{h^{\lambda_Z}}}\mathbb{E}\left[K\left(\frac{Z-z}{h}\right)\left(\psi(W,\hat{p},\hat{m}_0,\hat{m}_1)-\psi(W,p,m_0,m_1)\right)\right]\right\rvert
\end{align*}
and
\begin{align*}
\left\lvert\hat{f}(z)^{-1}\right\rvert\leq\sup_{z\in\mathcal{Z}}\left\lvert\hat{f}(z)^{-1}\right\rvert=\left\lvert\frac{1}{\inf_{z\in\mathcal{Z}}\hat{f}(z)}\right\rvert\leq\frac{1}{C}=O(1)
\end{align*}
by Assumption \ref{ass:kernel}.\\
Further, under the sample splitting procedure used
\begin{align*}
&\mathbb{E}\left[K\left(\frac{Z-z}{h}\right)\left(\psi(W,\hat{p},\hat{m}_0,\hat{m}_1)-\psi(W,p,m_0,m_1)\right)\right]\\
&=\mathbb{E}\left[K\left(\frac{Z-z}{h}\right)\left(\psi(W,\hat{p},\hat{m}_0,\hat{m}_1)-\psi(W,p,m_0,m_1)\right)|W_{i\in\mathcal{I}_l^c}\right]\\
& \leq \sup_{p^*\in\mathcal{P},m_0^*\in\mathcal{M}_0,m_1^*\in\mathcal{M}_1} \mathbb{E}\left[K\left(\frac{Z-z}{h}\right)\left(\psi(W,p^*,m^*_0,m^*_1)-\psi(W,p,m_0,m_1)\right)\right]
\end{align*}
Define the G\^{a}teaux derivative of the generic function $g$ in the direction $[p^*-p,m^*_0-m_0,m^*_1-m_1]$ by $\partial_{[p^*-p,m^*_0-m_0,m^*_1-m_1]} g$. Then using Taylor's expansion we can write
\begin{align*}
&\mathbb{E}\left[K\left(\frac{Z-z}{h}\right)\left(\psi(W,p^*,m^*_0,m^*_1)-\psi(W,p,m_0,m_1)\right)\right]\\
&=\partial_{[p^*-p,m^*_0-m_0,m^*_1-m_1]}\mathbb{E}\left[K\left(\frac{Z-z}{h}\right)\psi(W,p,m_0,m_1)\right]\\
&+\frac{1}{2}\partial^2_{[p^*-p,m^*_0-m_0,m^*_1-m_1]}\mathbb{E}\left[K\left(\frac{Z-z}{h}\right)\psi(W,p,m_0,m_1)\right]+...
\end{align*}
For the first order term we get
\begin{align*}
&\partial_{[p^*-p,m^*_0-m_0,m^*_1-m_1]}\mathbb{E}\left[K\left(\frac{Z-z}{h}\right)\psi(W,p,m_0,m_1)\right]\\
&=\mathbb{E}\Bigg[K\left(\frac{Z-z}{h}\right)\Bigg(-\left(\frac{D(Y-m_1(X))}{p(X)^2}+\frac{(1-D)(Y-m_0(X))}{(1-p(X))^2}\right)\left(p^*(X)-p(X)\right)\\
&+\left(\frac{(1-D)}{1-p(X)}-1\right)\left(m^*_0(X)-m_0(X)\right)+\left(1-\frac{D}{p(X)}\right)\left(m^*_1(X)-m_1(X)\right)\Bigg)\Bigg]\\
&=0
\end{align*}
by the Law of Iterated Expectation and using the fact that $Z\subseteq X$. For the second order term we get
\begin{align*}
&\frac{1}{2}\partial^2_{[p^*-p,m^*_0-m_0,m^*_1-m_1]}\mathbb{E}\left[K\left(\frac{Z-z}{h}\right)\psi(W,p,m_0,m_1)\right]\\
&=\mathbb{E}\Bigg[K\left(\frac{Z-z}{h}\right)\Bigg(\left(\frac{D(Y-m_1(X))}{p(X)^3}-\frac{(1-D)(Y-m_0(X))}{(1-p(X))^3}\right)\left(p^*(X)-p(X)\right)^2\\
&+\frac{1-D}{(1-p(X))^2}\left(p^*(X)-p(X)\right)\left(m^*_0(X)-m_0(X)\right)\\
&+\frac{D}{p(X)^2}\left(p^*(X)-p(X)\right)\left(m^*_1(X)-m_1(X)\right)\Bigg)\Bigg]\\
&=\mathbb{E}\Bigg[K\left(\frac{Z-z}{h}\right)\Bigg(\frac{1}{(1-p(X))}\left(p^*(X)-p(X)\right)\left(m^*_0(X)-m_0(X)\right)\\
&+\frac{1}{p(X)}\left(p^*(X)-p(X)\right)\left(m^*_1(X)-m_1(X)\right)\Bigg)\Bigg]\\
&\leq\left\lVert K(u)\right\rVert_{\infty}\times\Bigg\lVert\mathbb{E}\Bigg[\frac{1}{(1-p(X))}\left(p^*(X)-p(X)\right)\left(m^*_0(X)-m_0(X)\right)\\
&+\frac{1}{p(X)}\left(p^*(X)-p(X)\right)\left(m^*_1(X)-m_1(X)\right)\Bigg |Z\Bigg]\Bigg\rVert_1\\
&\leq C\left\lVert\frac{1}{(1-p(X))}\left(p^*(X)-p(X)\right)\left(m^*_0(X)-m_0(X)\right)+\frac{1}{p(X)}\left(p^*(X)-p(X)\right)\left(m^*_1(X)-m_1(X)\right)\right\rVert_1\\
&\leq C\times\left\lVert p^*(X)-p(X)\right\rVert_2\times\left(\left\lVert m^*_0(X)-m_0(X)\right\rVert_2+\left\lVert m^*_1(X)-m_1(X)\right\rVert_2\right)
\end{align*}
which follows from Hölder's and Jensen's inequality, $\left\lVert K(u)\right\rVert_{\infty}=O(1)$ in Assumption \ref{ass:kernel} and Assumption \ref{ass:cs}. All higher order terms can be shown to be dominated by the second order term under the boundedness Assumption \ref{ass:superror}. Therefore
\begin{align*}
\mathbb{E}\left[K\left(\frac{Z-z}{h}\right)\left(\psi(W,\hat{p},\hat{m}_0,\hat{m}_1)-\psi(W,p,m_0,m_1)\right)\right]=O(\epsilon_p\epsilon_{m_0} + \epsilon_p\epsilon_{m_1})
\end{align*}
and
\begin{align*}
\left\lvert \sqrt{Nh^{\lambda_Z}}iia\right\rvert=O\left(N^\frac{1}{2}h^{-\frac{1}{2}\lambda_Z}\times\left(\epsilon_p\epsilon_{m_0} + \epsilon_p\epsilon_{m_1}\right)\right).\\~\\
\end{align*}
\textit{Bounding iib}\\
We can write
\begin{align*}
\left\lvert \sqrt{Nh^{\lambda_Z}}iia\right\rvert=&\left\lvert\frac{\frac{1}{\sqrt{Nh^{\lambda_Z}}}\sum_{i=1}^N\left(K\left(\frac{z_i-z}{h}\right)\bar{\psi}_i-\mathbb{E}\left[K\left(\frac{Z-z}{h}\right)\bar{\psi}\right]\right)}{\frac{1}{Nh^{\lambda_Z}}\sum_{i=1}^NK\left(\frac{z_i-z}{h}\right)}\right\rvert\\
\leq & \frac{1}{\sqrt{h^{\lambda_Z}}}\left\lvert\hat{f}(z)^{-1}\right\rvert\left\lvert\frac{1}{\sqrt{N}}\sum_{i=1}^N\left(K\left(\frac{z_i-z}{h}\right)\bar{\psi}_i-\mathbb{E}\left[K\left(\frac{Z-z}{h}\right)\bar{\psi}\right]\right)\right\rvert\\
\leq & \frac{1}{\sqrt{h^{\lambda_Z}}}\frac{1}{C}\left\lvert\frac{1}{\sqrt{N}}\sum_{i=1}^N\left(K\left(\frac{z_i-z}{h}\right)\bar{\psi}_i-\mathbb{E}\left[K\left(\frac{Z-z}{h}\right)\bar{\psi}\right]\right)\right\rvert
\end{align*}
which follows again from Assumption \ref{ass:kernel}. The convergence of the last factor term remains to show. Since $L$ is a fixed integer that is independent of $N$, it suffices to show that for any $l\in[L]$ the term converges. More formally
\begin{align*}
\left\lvert\frac{1}{\sqrt{N}}\sum_{i=1}^N\left(K\left(\frac{z_i-z}{h}\right)\bar{\psi}_i-\mathbb{E}\left[K\left(\frac{Z-z}{h}\right)\bar{\psi}\right]\right)\right\rvert\leq\max_{l\in[L]}\left\lvert\frac{1}{\sqrt{nL}}\sum_{i\in\mathcal{I}_l}^n\left(K\left(\frac{z_i-z}{h}\right)\bar{\psi}_i-\mathbb{E}\left[K\left(\frac{Z-z}{h}\right)\bar{\psi}\right]\right)\right\rvert
\end{align*}
where $\mathcal{I}_l$ is the set of observation in subsample $l$ and $\mathcal{I}_l^c$ is the set of observations not in subsample $l$.\\
Under the sample splitting procedure we have
\begin{align*}
&\mathbb{E}\left[\left\lvert\frac{1}{\sqrt{n}}\sum_{i\in\mathcal{I}_l}^n\left(K\left(\frac{z_i-z}{h}\right)\bar{\psi}_i-\mathbb{E}\left[K\left(\frac{Z-z}{h}\right)\bar{\psi}\right]\right)\right\rvert^2\right]\\
&=\mathbb{E}\left[\left\lvert\frac{1}{\sqrt{n}}\sum_{i\in\mathcal{I}_l}^n\left(K\left(\frac{z_i-z}{h}\right)\bar{\psi}_i-\mathbb{E}\left[K\left(\frac{Z-z}{h}\right)\bar{\psi}\right]\right)\right\rvert^2\Bigg|W_{i\in\mathcal{I}_l^c}\right]\\
&\leq \sup_{p^*\in\mathcal{P},m_0^*\in\mathcal{M}_0,m_1^*\in\mathcal{M}_1}\mathbb{E}\left[\left\lvert K\left(\frac{z_i-z}{h}\right)\left(\psi(W,p^*,m^*_0,m^*_1)-\psi(W,p,m_0,m_1)\right)\right\rvert^2\right]\\
&\leq\sup_{u}\left\lvert K(u)^2\right\rvert\times\sup_{p^*\in\mathcal{P},m_0^*\in\mathcal{M}_0,m_1^*\in\mathcal{M}_1}\left\lVert\mathbb{E}\left[\left(\psi(W,p^*,m^*_0,m^*_1)-\psi(W,p,m_0,m_1)\right)^2|Z\right]\right\rVert_1\\
&\leq\left\lVert K(u)\right\rVert_\infty^2\sup_{p^*\in\mathcal{P},m_0^*\in\mathcal{M}_0,m_1^*\in\mathcal{M}_1}\mathbb{E}\left[\left\lvert\psi(W,p^*,m^*_0,m^*_1)-\psi(W,p,m_0,m_1)\right\rvert^2\right]\\
&\leq C \sup_{p^*\in\mathcal{P},m_0^*\in\mathcal{M}_0,m_1^*\in\mathcal{M}_1}\mathbb{E}\left[\left\lvert\psi(W,p^*,m^*_0,m^*_1)-\psi(W,p,m_0,m_1)\right\rvert^2\right]
\end{align*}
by Hölder's inequality, Jensen's inequality and Assumption \ref{ass:kernel}. Now
\begin{align*}
&\sup_{p^*\in\mathcal{P},m_0^*\in\mathcal{M}_0,m_1^*\in\mathcal{M}_1}\left(\mathbb{E}\left[\left\lvert\psi(W,p^*,m^*_0,m^*_1)-\psi(W,p,m_0,m_1)\right\rvert^2\right]\right)^{\frac{1}{2}}\\
&=\sup_{p^*\in\mathcal{P},m_0^*\in\mathcal{M}_0,m_1^*\in\mathcal{M}_1}\Bigg(\mathbb{E}\Bigg[\Big\lvert\frac{D(Y-m^*_1(X))}{p^*(X)}-\frac{(1-D)(Y-m^*_0(X))}{1-p^*(X)}+m^*_1(X)-m^*_0(X)-\frac{D(Y-m_1(X))}{p(X)}\\
&+\frac{(1-D)(Y-m_0(X))}{1-p(X)}-m_1(X)+m_0(X)\Big\rvert^2\Bigg]\Bigg)^{\frac{1}{2}}\\
&\leq\sup_{p^*\in\mathcal{P},m_0^*\in\mathcal{M}_0,m_1^*\in\mathcal{M}_1}\left\lVert m^*_1(X)-m_1(X)\right\rVert_2+\sup_{p^*\in\mathcal{P},m_0^*\in\mathcal{M}_0,m_1^*\in\mathcal{M}_1}\left\lVert m^*_0(X)-m_0(X)\right\rVert_2\\
&+\sup_{p^*\in\mathcal{P},m_0^*\in\mathcal{M}_0,m_1^*\in\mathcal{M}_1}\left\lVert\frac{D(Y-m^*_1(X))}{p^*(X)}-\frac{D(Y-m_1(X))}{p(X)}\right\rVert_2\\
&+\sup_{p^*\in\mathcal{P},m_0^*\in\mathcal{M}_0,m_1^*\in\mathcal{M}_1}\left\lVert\frac{(1-D)(Y-m^*_0(X))}{1-p^*(X)}-\frac{(1-D)(Y-m_0(X))}{1-p(X)}\right\rVert_2
\end{align*}
and by defining $U=DY-m_1(X)$
\begin{align*}
\left\lVert\frac{D(Y-m^*_1(X))}{p^*(X)}-\frac{D(Y-m_1(X))}{p(X)}\right\rVert_2=&\left\lVert\frac{1}{p(X)p^*(X)}\left(D(Y-m^*_1(X))p(X)-D(Y-m_1(X))p^*(X)\right)\right\rVert_2\\
\leq & c^{-2}\left\lVert D(Y-m^*_1(X))p(X)-D(Y-m_1(X))p^*(X)\right\rVert_2\\
=& c^{-2}\left\lVert p(X)(m_1(X)-m^*_1(X))+U(p(X)-p^*(X))\right\rVert_2\\
\leq & c^{-2}\left\lVert m_1(X)-m^*_1(X)\right\rVert_2+ c^{-2}\left\lVert U(p(X)-p^*(X))\right\rVert_2.
\end{align*}
Since
\begin{align*}
\left\lVert U(p(X)-p^*(X))\right\rVert_2=&\sqrt{\mathbb{E}\left[\left(U(p(X)-p^*(X))\right)^2\right]}\\
=&\sqrt{\mathbb{E}\left[\mathbb{E}\left[U^2|X\right](p(X)-p^*(X))^2\right]}
\end{align*}
and a similar argument for the other term by Assumption \ref{ass:error} we get
\begin{align*}
&\sup_{p^*\in\mathcal{P},m_0^*\in\mathcal{M}_0,m_1^*\in\mathcal{M}_1}\left\lVert m^*_1(X)-m_1(X)\right\rVert_2=\epsilon_{m_1}\\
&\sup_{p^*\in\mathcal{P},m_0^*\in\mathcal{M}_0,m_1^*\in\mathcal{M}_1}\left\lVert m^*_0(X)-m_0(X)\right\rVert_2=\epsilon_{m_0}\\
&\sup_{p^*\in\mathcal{P},m_0^*\in\mathcal{M}_0,m_1^*\in\mathcal{M}_1}\left\lVert\frac{D(Y-m^*_1(X))}{p^*(X)}-\frac{D(Y-m_1(X))}{p(X)}\right\rVert_2=\epsilon_{m_1}+\epsilon_p\\
&\sup_{p^*\in\mathcal{P},m_0^*\in\mathcal{M}_0,m_1^*\in\mathcal{M}_1}\left\lVert\frac{(1-D)(Y-m^*_0(X))}{1-p^*(X)}-\frac{(1-D)(Y-m_0(X))}{1-p(X)}\right\rVert_2=\epsilon_{m_0}+\epsilon_p.
\end{align*}
It follows that
\begin{align*}
\sup_{p^*\in\mathcal{P},m_0^*\in\mathcal{M}_0,m_1^*\in\mathcal{M}_1}\mathbb{E}\left[\left\lvert\psi(W,p^*,m^*_0,m^*_1)-\psi(W,p,m_0,m_1)\right\rvert^2\right]\leq \max(\epsilon_p,\epsilon_{m_0},\epsilon_{m_1})^2=\epsilon_{\max}^2.
\end{align*}
By Markov's inequality and the fact that if $L$ is a constant independent of $N$ it follows that
\begin{align*}
\left\lvert\frac{1}{\sqrt{N}}\sum_{i=1}^N\left(K\left(\frac{z_i-z}{h}\right)\bar{\psi}_i-\mathbb{E}\left[K\left(\frac{Z-z}{h}\right)\bar{\psi}\right]\right)\right\rvert\leq C\times\epsilon_{\max}
\end{align*}
and therefore
\begin{align*}
\left\lvert\sqrt{Nh^{\lambda_Z}}iia\right\rvert=O(h^{-\frac{1}{2}\lambda_Z}\epsilon_{\max}).\\
\end{align*}
Collecting terms, we can write
\begin{align*}
\sqrt{Nh^{\lambda_Z}}\left(\hat{\tau}-\tau\right)=&\frac{\frac{1}{\sqrt{Nh^{\lambda_Z}}}\sum_{i=1}^NK\left(\frac{z_i-z}{h}\right)\left(\psi(W_i,p,m_0,m_1)-\tau\right)}{\frac{1}{Nh^{\lambda_Z}}\sum_{i=1}^NK\left(\frac{z_i-z}{h}\right)}\\
&+O\left(N^\frac{1}{2}h^{-\frac{1}{2}\lambda_Z}\times\left(\epsilon_p\epsilon_{m_0} + \epsilon_p\epsilon_{m_1}\right)+h^{-\frac{1}{2}\lambda_Z}\epsilon_{\max}\right).
\end{align*}
Under the convergence conditions in Assumption \ref{ass:cate} the first claim of the theorem is verified.
\subsubsection*{Asymptotic normality}
Notice that under the standard conditions provided in Assumptions \ref{ass:kernel} and \ref{ass:cate} on the nonparametric regression (see for example \cite[chapter 2]{Pagan_Ullah_1999})
\begin{align*}
\frac{1}{Nh^{\lambda_Z}}\sum_{i=1}^NK\left(\frac{z_i-z}{h}\right)\rightarrow_p f(z).
\end{align*}
Therefore, we can rewrite the influence function as
\begin{align*}
\sqrt{Nh^{\lambda_Z}}\left(\hat{\tau}-\tau\right)=&\frac{1}{\sqrt{Nh^{\lambda_Z}}}\frac{1}{f(z)}\sum_{i=1}^NK\left(\frac{z_i-z}{h}\right)\left(\psi(W_i,p,m_0,m_1)-\tau\right)+o(1)\\
=&\underbrace{\frac{1}{\sqrt{Nh^{\lambda_Z}}}\frac{1}{f(z)}\sum_{i=1}^NK\left(\frac{z_i-z}{h}\right)\left(\psi(W_i,p,m_0,m_1)-\mathbb{E}\left[\psi(W_i,p,m_0,m_1)|Z=z_i\right]\right)}_{ia}\\
&+\underbrace{\frac{1}{\sqrt{Nh^{\lambda_Z}}}\frac{1}{f(z)}\sum_{i=1}^NK\left(\frac{z_i-z}{h}\right)\left(\mathbb{E}\left[\psi(W_i,p,m_0,m_1)|Z=z_i\right]-\tau\right)}_{ib}\\
&+o(1).
\end{align*}
The second term is the bias of the nonparametric regression estimator scaled with the convergence rate. Thus, Assumption \ref{ass:cate} implies $ib=O(N^{\frac{1}{2}}h^{\frac{1}{2}\lambda_Z}h^r)=o(1)$. Under the usual assumptions on the existence of higher order moments in Assumption \ref{ass:kernel}, we can apply the Lyapunov Central Limit Theorem on $ia$ as in \textcite[chapter 3.4]{Pagan_Ullah_1999}. Then
\begin{align*}
\sqrt{Nh^{\lambda_Z}}\left(\hat{\tau}-\tau\right)\rightarrow_d N\left(0,\frac{\int K(u)^2du\times\mathbb{E}\left[\left(\psi(W_i,p,m_0,m_1)-\tau\right)^2|Z=z\right]}{f(z)}\right).
\end{align*}
\begin{flushright}
\textit{q.e.d.}
\end{flushright}
\subsection{Proof of Theorem \ref{thm:ate}}
Similar to the proof in Theorem \ref{thm:cate} we can write
\begin{align*}
\hat{\theta}-\theta=&\frac{1}{N}\sum_{i=1}^N\sum_{j=1}^N\frac{\frac{1}{Nh^{\lambda_Z}}K\left(\frac{z_j-z_i}{h}\right)\psi\left(W_j,\hat{p},\hat{m}_0,\hat{m}_1\right)}{\frac{1}{Nh^{\lambda_Z}}\sum_{j=1}^NK\left(\frac{z_j-z_i}{h}\right)}-\theta\\
=&\underbrace{\frac{1}{N}\sum_{i=1}^N\sum_{j=1}^N\frac{\frac{1}{Nh^{\lambda_Z}}K\left(\frac{z_j-z_i}{h}\right)\psi\left(W_j,p,m_0,m_1\right)}{\frac{1}{Nh^{\lambda_Z}}\sum_{j=1}^NK\left(\frac{z_j-z_i}{h}\right)}-\theta}_{i}\\
&+\underbrace{\frac{1}{N}\sum_{i=1}^N\sum_{j=1}^N\frac{\frac{1}{Nh^{\lambda_Z}}K\left(\frac{z_j-z_i}{h}\right)\left(\psi\left(W_j,\hat{p},\hat{m}_0,\hat{m}_1\right)-\psi\left(W_j,p,m_0,m_1\right)\right)}{\frac{1}{Nh^{\lambda_Z}}\sum_{j=1}^NK\left(\frac{z_j-z_i}{h}\right)}}_{ii}.
\end{align*}
\textit{Bounding ii}\\
Using the notation from the proof of Theorem \ref{thm:cate} again leads to
\begin{align*}
ii=&\underbrace{\frac{1}{N}\sum_{i=1}^N\sum_{j=1}^N\frac{\frac{1}{Nh^{\lambda_Z}}\mathbb{E}\left[K\left(\frac{Z-z_i}{h}\right)\bar{\psi}\right]}{\frac{1}{Nh^{\lambda_Z}}\sum_{j=1}^NK\left(\frac{z_j-z_i}{h}\right)}}_{iia}\\
&+\underbrace{\frac{1}{N}\sum_{i=1}^N\sum_{j=1}^N\frac{\frac{1}{Nh^{\lambda_Z}}\left(K\left(\frac{z_j-z_i}{h}\right)\bar{\psi}_j-\mathbb{E}\left[K\left(\frac{Z-z_i}{h}\right)\bar{\psi}\right]\right)}{\frac{1}{Nh^{\lambda_Z}}\sum_{j=1}^NK\left(\frac{z_j-z_i}{h}\right)}}_{iib}.
\end{align*}
Then
\begin{align*}
\left\lvert iia\right\rvert=&\left\lvert\frac{1}{N}\sum_{i=1}^N\frac{\frac{1}{h^{\lambda_Z}}\mathbb{E}\left[K\left(\frac{Z-z_i}{h}\right)\bar{\psi}\right]}{\frac{1}{Nh^{\lambda_Z}}\sum_{j=1}^NK\left(\frac{z_j-z_i}{h}\right)}\right\rvert\\
&\leq\sup_{i}\left\lvert\frac{\frac{1}{h^{\lambda_Z}}\mathbb{E}\left[K\left(\frac{Z-z_i}{h}\right)\bar{\psi}\right]}{\frac{1}{Nh^{\lambda_Z}}\sum_{j=1}^NK\left(\frac{z_j-z_i}{h}\right)}\right\rvert\\
&\leq\frac{1}{h^{\lambda_Z}}\sup_{z\in\mathcal{Z}}\left\lvert \hat{f}(z)^{-1}\right\rvert\times\sup_i\left\lvert\mathbb{E}\left[K\left(\frac{Z-z_i}{h}\right)\bar{\psi}\right]\right\rvert.
\end{align*}
By the same steps as in the proof of Theorem \ref{thm:cate} we obtain
\begin{align*}
\mathbb{E}\left[K\left(\frac{Z-z}{h}\right)\left(\psi(W,\hat{p},\hat{m}_0,\hat{m}_1)-\psi(W,p,m_0,m_1)\right)\right]=O(\epsilon_p\epsilon_{m_0} + \epsilon_p\epsilon_{m_1})
\end{align*}
and therefore
\begin{align*}
\left\lvert\sqrt{N}iia\right\rvert=O\left(\frac{\sqrt{N}}{h^{\lambda_Z}}\epsilon_p\times(\epsilon_{m_0}+\epsilon_{m_1})\right).
\end{align*}
Also for $iib$ we find that
\begin{align*}
\left\lvert\sqrt{N}iib\right\rvert=&\left\lvert\frac{1}{N}\sum_{i=1}^N\sum_{j=1}^N\frac{\frac{1}{\sqrt{N}h^{\lambda_Z}}\left(K\left(\frac{z_j-z_i}{h}\right)\bar{\psi}_i-\mathbb{E}\left[K\left(\frac{Z-z_i}{h}\right)\bar{\psi}\right]\right)}{\frac{1}{Nh^{\lambda_Z}}\sum_{j=1}^NK\left(\frac{z_j-z_i}{h}\right)}\right\rvert\\
&\leq\frac{1}{h^{\lambda_Z}}\sup_{z\in\mathcal{Z}}\left\lvert\hat{f}(z)^{-1}\right\rvert\times\sup_i\left\lvert\frac{1}{\sqrt{N}}\sum_{j=1}^N K\left(\frac{z_j-z_i}{h}\right)\bar{\psi}_i-\mathbb{E}\left[K\left(\frac{Z-z_i}{h}\right)\bar{\psi}\right]\right\rvert\\
&\leq C\frac{\epsilon_{\max}}{h^{\lambda_Z}}.
\end{align*}
Hence, for the overall term we have
\begin{align*}
\left\lvert \sqrt{N}ii\right\rvert=O\left(\frac{\sqrt{N}}{h^{\lambda_Z}}\epsilon_p\times(\epsilon_{m_0}+\epsilon_{m_1})+\frac{\epsilon_{\max}}{h^{\lambda_Z}}\right)=o(1),
\end{align*}
under the coupled convergence conditions of Assumption \ref{ass:ate}.\\~\\
\textit{Bounding i}\\
For $i$ notice that
\begin{align*}
\sqrt{N}i=\sqrt{N}\left(\tilde{\theta}-\theta\right),
\end{align*}
where 
\begin{align*}
\tilde{\theta}=\frac{1}{N}\sum_{i=1}^N\sum_{j=1}^N\frac{\frac{1}{Nh^{\lambda_Z}}K\left(\frac{z_j-z_i}{h}\right)\psi\left(W_j,p,m_0,m_1\right)}{\frac{1}{Nh^{\lambda_Z}}\sum_{j=1}^NK\left(\frac{z_j-z_i}{h}\right)}.
\end{align*}
Thus, term $i$ gives the contribution of estimating the nonparametric projection of the vector $\psi=\psi(W,p,m_0,m_1)$ with population nuisance parameters on $Z\in\mathcal{Z}$. To derive the influence function of the estimator $\tilde{\theta}$, we follow \posscite{Newey_1994} Proposition 4 which holds under the condition that the first stage nonparametric estimator is bounded by any norm (and some further regularity conditions). For example, using Assumption \ref{ass:ate} we have
\begin{align*}
N^{\frac{1}{4}}\left\lVert\sum_{j=1}^N\frac{\frac{1}{Nh^{\lambda_Z}}K\left(\frac{z_j-z}{h}\right)\psi\left(W_j,p,m_0,m_1\right)}{\frac{1}{Nh^{\lambda_Z}}\sum_{j=1}^NK\left(\frac{z_j-z}{h}\right)}-\tau(z)\right\rVert_2=o(1),
\end{align*}
such that Assumption 5.1 in \textcite{Newey_1994} is satisfied for the $L_2$ norm. In particular, we notice that the influence function $\phi$ is composed of the moment condition and an adjustment term. The moment condition of the problem is given by
\begin{align*}
\mathbb{E}\left[\tilde{\tau}(Z)-\theta\right]=0
\end{align*}
with $\tilde{\tau}(Z)=\mathbb{E}\left[\psi|Z\right]$.\\
Denote the general family of distributions of $W=(Y,D,X,Z)$ as $\mathcal{F}=\{F(W)\}$. Further, denote $F_\beta(W)\in\mathcal{F}$ a subfamily of $\mathcal{F}$ that is a path in $\mathcal{F}$ indexed by $\beta$. Also let $F_0$ be the true distribution of $W$. Accordingly, $W$ realizes with density $f_\beta(W)$ when $\beta=0$. Additionally, define $\mathbb{E}_\beta\left[g(W)\right]=\int g(W)f_\beta(W)dw$ for the generic function $g(\cdot)$ and $\tilde{\tau}(Z,\beta)=\mathbb{E}_\beta\left[\psi|Z\right]$. Following the steps of Proposition 4 in \textcite{Newey_1994} indicates that one should evaluate the derivative
\begin{align*}
\frac{\partial}{\partial\beta}\mathbb{E}\left[\tilde{\tau}(Z,\beta)\right]
\end{align*}
at $\beta=0$. By the Chain Rule we have
\begin{align*}
\frac{\partial}{\partial\beta}\mathbb{E}_\beta\left[\tilde{\tau}(Z,\beta)\right]=\frac{\partial}{\partial\beta}\mathbb{E}_\beta\left[\tilde{\tau}(Z)\right]+\frac{\partial}{\partial\beta}\mathbb{E}\left[\tilde{\tau}(Z,\beta)\right]
\end{align*}
at $\beta=0$. Furthermore, for any $\bar{\tau}(Z)$ and for some path $\beta$ we have the mean-square projection optimization problem
\begin{align*}
\tilde{\tau}(Z,\beta)=\text{arg}\,\max_{\bar{\tau}}\mathbb{E}_\beta\left[\left(\frac{D(Y-m_1(X))}{p(X)}-\frac{(1-D)(Y-m_0(X))}{1-p(X)}+m_1(X)-m_0(X)-\bar{\tau}(Z)\right)^2\right],
\end{align*}
giving the first order condition
\begin{align*}
\mathbb{E}_\beta\left[\frac{D(Y-m_1(X))}{p(X)}-\frac{(1-D)(Y-m_0(X))}{1-p(X)}+m_1(X)-m_0(X)-\tilde{\tau}(Z,\beta)\right]=0.
\end{align*}
Define $S(W)=\frac{\partial}{\partial\beta}\ln f_\beta(W)$ at $\beta=0$. Then combining the two previous result gives
\begin{align*}
\frac{\partial}{\partial\beta}\mathbb{E}\left[\tilde{\tau}(Z,\beta)\right]=&\frac{\partial}{\partial\beta}\mathbb{E}_\beta\left[\tilde{\tau}(Z,\beta)\right]-\frac{\partial}{\partial\beta}\mathbb{E}_\beta\left[\tilde{\tau}(Z)\right]\\
=&\frac{\partial}{\partial\beta}\mathbb{E}_\beta\left[\frac{D(Y-m_1(X))}{p(X)}-\frac{(1-D)(Y-m_0(X))}{1-p(X)}+m_1(X)-m_0(X)-\tilde{\tau}(Z)\right]\\
=&\mathbb{E}\left[\left(\frac{D(Y-m_1(X))}{p(X)}-\frac{(1-D)(Y-m_0(X))}{1-p(X)}+m_1(X)-m_0(X)-\tilde{\tau}(Z)\right)S(W)\right]
\end{align*}
at $\beta=0$. It follows that the adjustment term is given by $\psi(W,p,m_0,m_1)-\tilde{\tau}(Z)$ and the influence function has the form
\begin{align*}
\phi&=\tilde{\tau}(Z)-\theta+\psi(W,p,m_0,m_1)-\tilde{\tau}(Z)\\
&=\frac{D(Y-m_1(X))}{p(X)}-\frac{(1-D)(Y-m_0(X))}{1-p(X)}+m_1(X)-m_0(X)-\theta.
\end{align*}
Hence, combining the results for terms $i$ and $ii$ gives
\begin{align*}
\sqrt{N}(\hat{\theta}-\theta)=\frac{1}{\sqrt{N}}\sum_{i=1}^N\left(\frac{d_i(y_i-m_1(x_i))}{p(x_i)}-\frac{(1-d_i)(y_i-m_0(x_i))}{1-p(x_i)}+m_1(x_i)-m_0(x_i)-\theta\right)+o(1)
\end{align*}
such that by the Central Limit Theorem we obtain
\begin{align*}
\sqrt{N}(\hat{\theta}-\theta)\rightarrow_d N\left(0,\mathbb{E}\left[\frac{Var(Y|D=1,X)}{p(X)}+\frac{Var(Y|D=0,X)}{1-p(X)}+(m_1(X)-m_0(X)-\theta)^2\right]\right).
\end{align*}
\begin{flushright}
\textit{q.e.d.}
\end{flushright}
\newpage
\section{Details on the bandwidth ranges}\label{app:range}
Using the notation implied by Definition \ref{def:rates}, notice that the convergence conditions in Assumption \ref{ass:cate} imply the following system of inequalities.
\begin{align*}
\frac{1}{2}\lambda_Z\delta_h-\delta_{\epsilon_{\max}}&<0\\
\frac{1}{2}+\frac{1}{2}\lambda_Z\delta_h-(\delta_p+\delta_{m_d})&<0\\
\frac{1}{2}-\frac{1}{2}\lambda_Z\delta_h-r\delta_h&<0\\
-\delta_h&<0\\
1-\delta_h\lambda_Z&>0
\end{align*}
The third and fourth inequality imply $\delta_h>\frac{1}{\lambda_Z+2r}>0$. Further, the other inequalities imply $\delta_h<\frac{2(\delta_p+\delta_{m_d})-1}{\lambda_Z}<\frac{2\delta_{\epsilon_{\max}}}{\lambda_Z}<\frac{1}{\lambda_Z}$. It therefore follows that the possible range of the bandwidth can be described by $\frac{1}{\lambda_Z+2r}<\delta_h<\frac{2(\delta_p+\delta_{m_d})-1}{\lambda_Z}$.\\
The range for ATE follows similarly from the convergence conditions in Assumption \ref{ass:ate}.
\newpage
\section{Details on the empirical example}\label{app:data}
\subsection{Covariates in the dataset}
\begin{table}[h]
\centering
\caption{Description of covariates in the dataset}
\label{tab:ressem}
\begin{threeparttable}
\begin{tabular}{l|llllc}
  \toprule
variable & description & newly created & smalldata & alldata & mean\\ 
  \midrule
Y & infant birth weight in grams & no & yes & yes & 3361.68\\
D & =1 if mother smoked during pregnancy & no & yes & yes & 0.19\\
mmarried & =1 if mother is married & no & no & yes & 0.70\\
mhisp & =1 if mother is hispanic & no & yes & yes & 0.03\\
fhisp & =1 if father is hispanic & no & no & yes & 0.04\\
foreign & =1 if mother born abroad & no & no & yes & 0.05\\
alcohol & =1 if alcohol consumed during pregnancy & no & yes & yes & 0.03\\
deadkids & =1 if previous birth were newborn died & no & yes & yes & 0.26\\
mage & mother's age & no & yes & yes & 26.50\\
medu & mother's educational attainment & no & yes & yes & 12.69\\
fage & father's age & no & no & yes & 27.27\\
fedu & father's educational attainment & no & no & yes & 12.31\\
nprenatal & number of prenatal care visits & no & yes & yes & 10.76\\
mrace & =1 if mother is white & no & yes & yes & 0.84\\
frace & =1 if father is white & no & no & yes & 0.81\\
prenatal1 & =1 if first prenatal visit in first trimester & no & yes & yes & 0.80\\
prenatal2 & =1 if first prenatal visit in second trimester & yes & no & yes & 0.15\\
prenatal3 & =1 if first prenatal visit in third trimester & yes & no & yes & 0.05\\
order1 & =1 if first infant & yes & yes & yes & 0.44\\
order2 & =1 if second infant & yes & no & yes & 0.34\\
order3 & =1 if $j^{th}$ infant with $j\geq 3$ & yes & no & yes & 0.22\\
birthmonth1 & =1 if birth in January & yes & no & yes & 0.07\\
birthmonth2 & =1 if birth in February & yes & no & yes & 0.08\\
birthmonth3 & =1 if birth in March & yes & no & yes & 0.08\\
birthmonth4 & =1 if birth in April & yes & no & yes & 0.08\\
birthmonth5 & =1 if birth in May & yes & no & yes & 0.08\\
birthmonth6 & =1 if birth in June & yes & no & yes & 0.10\\
birthmonth7 & =1 if birth in July & yes & no & yes & 0.08\\
birthmonth8 & =1 if birth in August & yes & no & yes & 0.10\\
birthmonth9 & =1 if birth in September & yes & no & yes & 0.09\\
birthmonth10 & =1 if birth in October & yes & no & yes & 0.08\\
birthmonth11 & =1 if birth in November & yes & no & yes & 0.08\\
birthmonth12 & =1 if birth in December & yes & no & yes & 0.08\\
\bottomrule \bottomrule
\end{tabular}
\begin{tablenotes}
\small 
\item Sample with $N=4642$ observations with $864$ treated and $3778$ non-treated. `smalldata' indicates that the variable was also used in \textcite{Lee_Okui_Whang_2016}. `newly created' indicates that the variable was additionally created from the original dataset by the authors. `alldata' contains the specification used for the estimation results in Section \ref{sec:example}.
\end{tablenotes}
\end{threeparttable}
\end{table}
\newpage
\subsection{Additional sensitivity analysis}
\begin{figure}[h]
\centering
\caption{Sensitivity to kernel order (age)}
\begin{subfigure}{0.49\textwidth}
\caption{Fourth order kernel function}
\includegraphics[width=\textwidth]{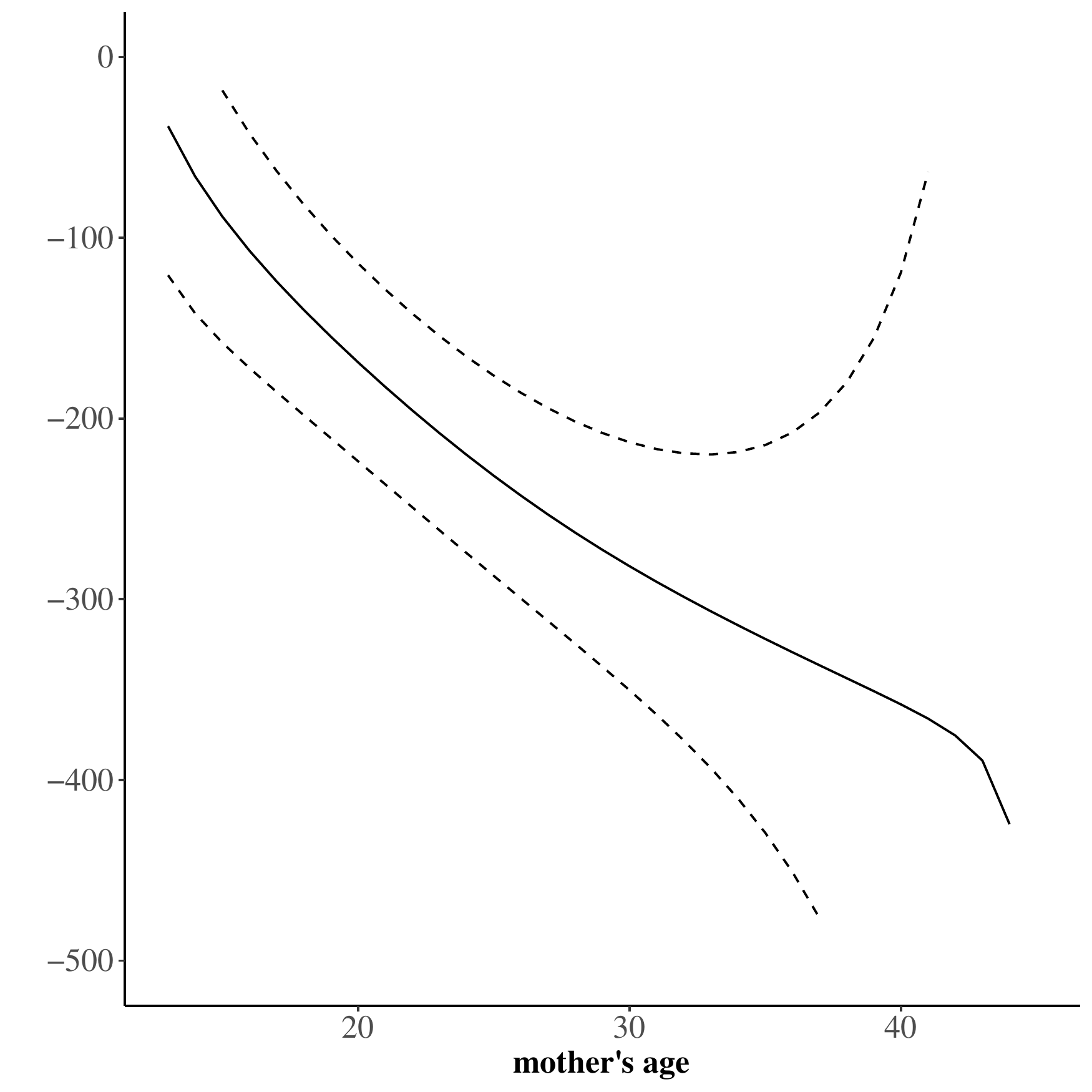}
\end{subfigure}
\begin{subfigure}{0.49\textwidth}
\caption{Sixth order kernel function}
\includegraphics[width=\textwidth]{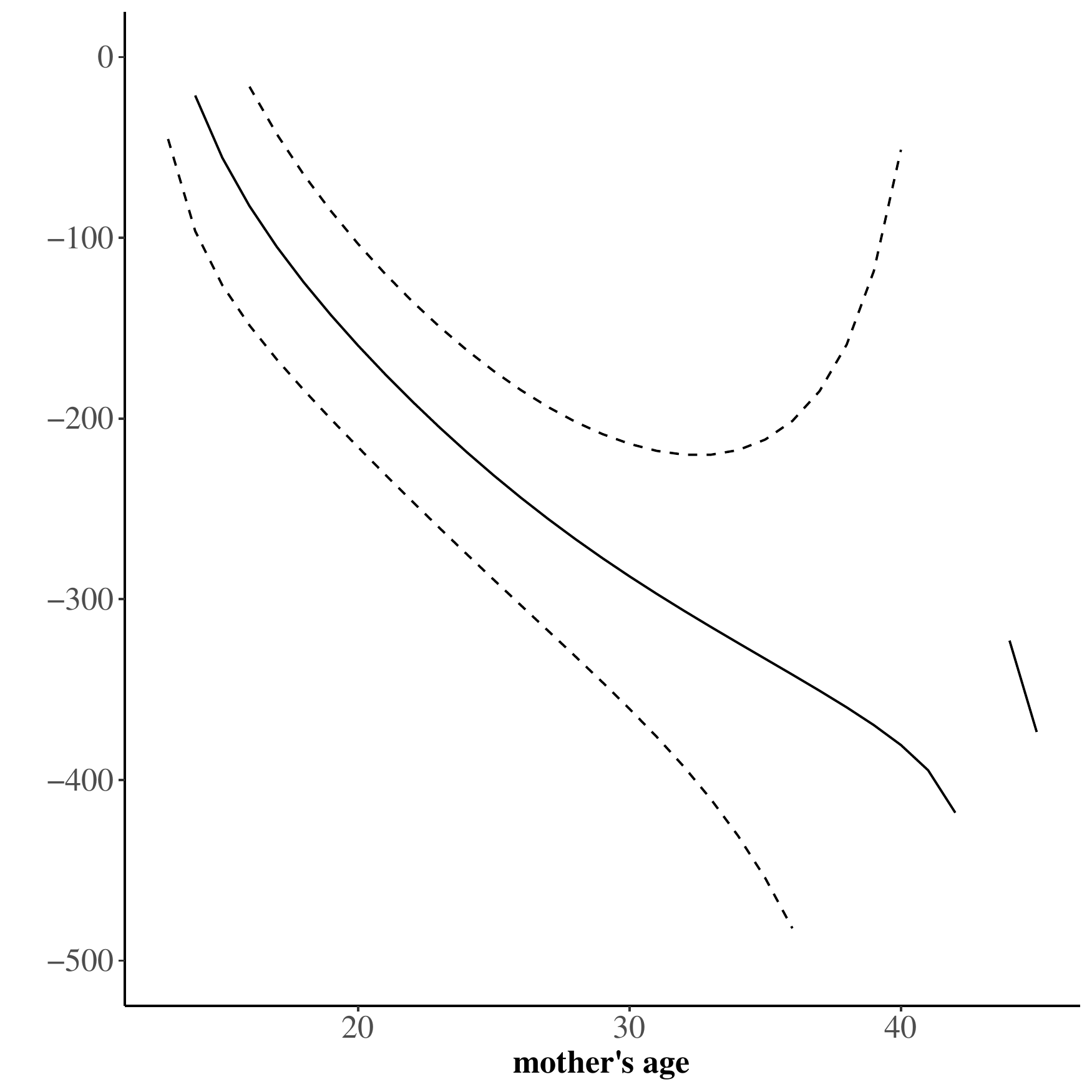}
\end{subfigure}
\label{fig:aipwkernel}
\floatfoot{Results were obtained as described in Procedure 1 with a fourth and sixth order Gaussian kernel function and $0.9\times$LOOCV bandwidth choice. Nuisance parameters were estimated using an ensemble learner comprising Lasso, Elastic Net, Ridge and Random Forest. For Lasso, Ridge and Elastic Net the penalty term was chosen such that the cross-validation criterion was minimized. The ensemble weights were chosen by minimizing out-of-sample MSE. Asymptotic confidence bands are at the $95\%$ level.}
\end{figure}
\begin{figure}[h]
\centering
\caption{Sensitivity to bandwidth choice (care visits)}
\begin{subfigure}{0.3\textwidth}
\caption{$0.5$ $\times$ CV choice}
\includegraphics[width=\textwidth]{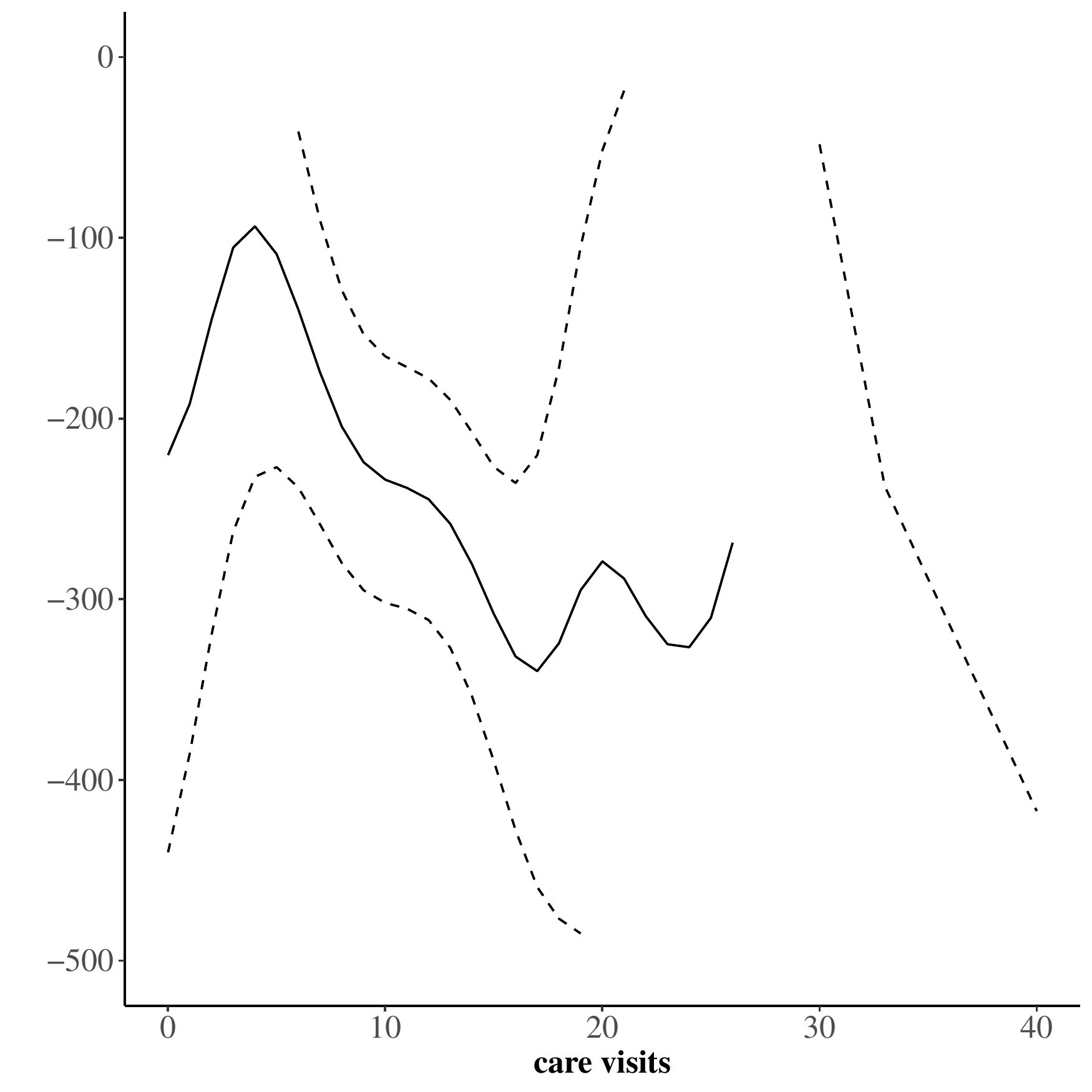}
\end{subfigure}
\begin{subfigure}{0.3\textwidth}
\caption{$0.7$ $\times$ CV choice}
\includegraphics[width=\textwidth]{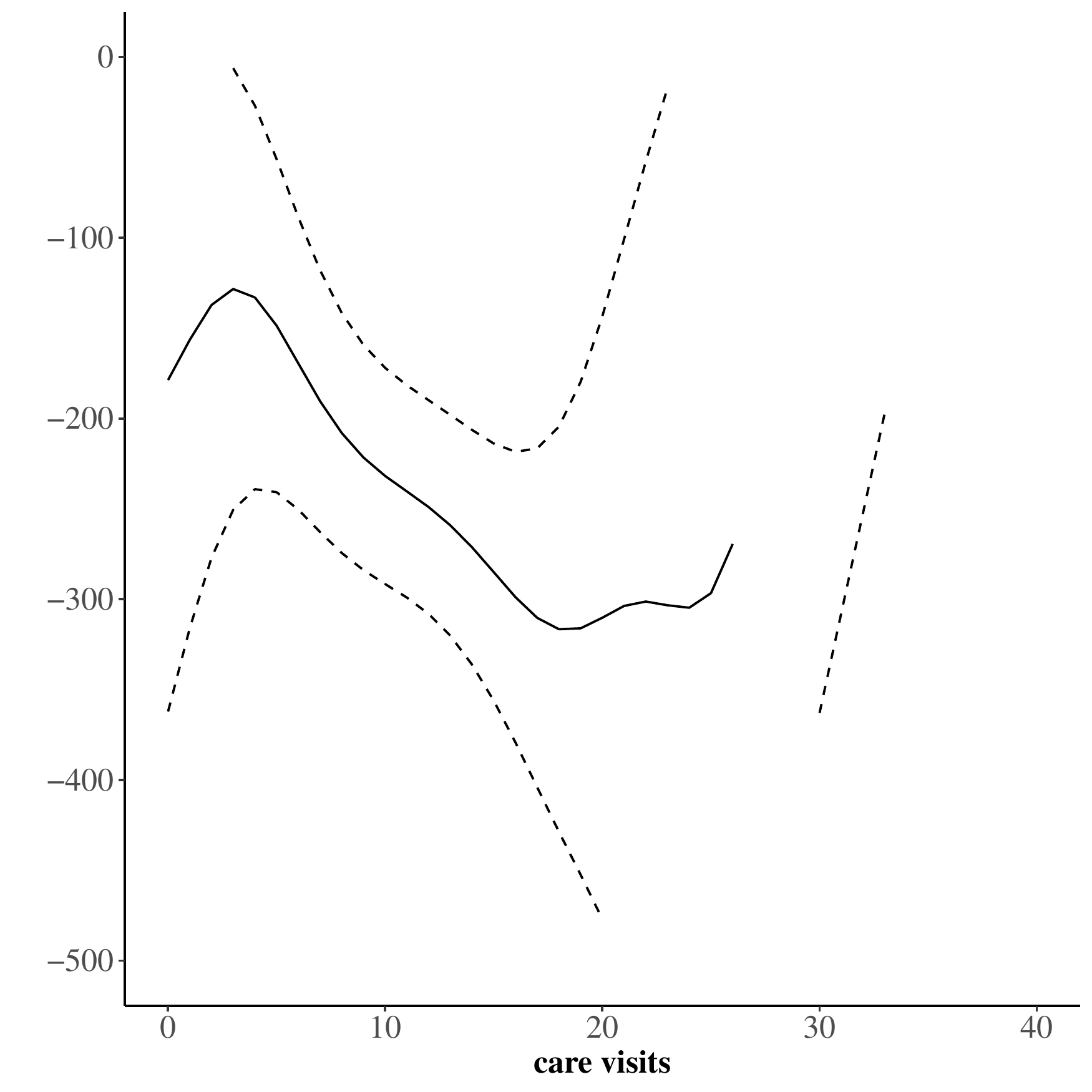}
\end{subfigure}
\begin{subfigure}{0.3\textwidth}
\caption{$0.8$ $\times$ CV choice}
\includegraphics[width=\textwidth]{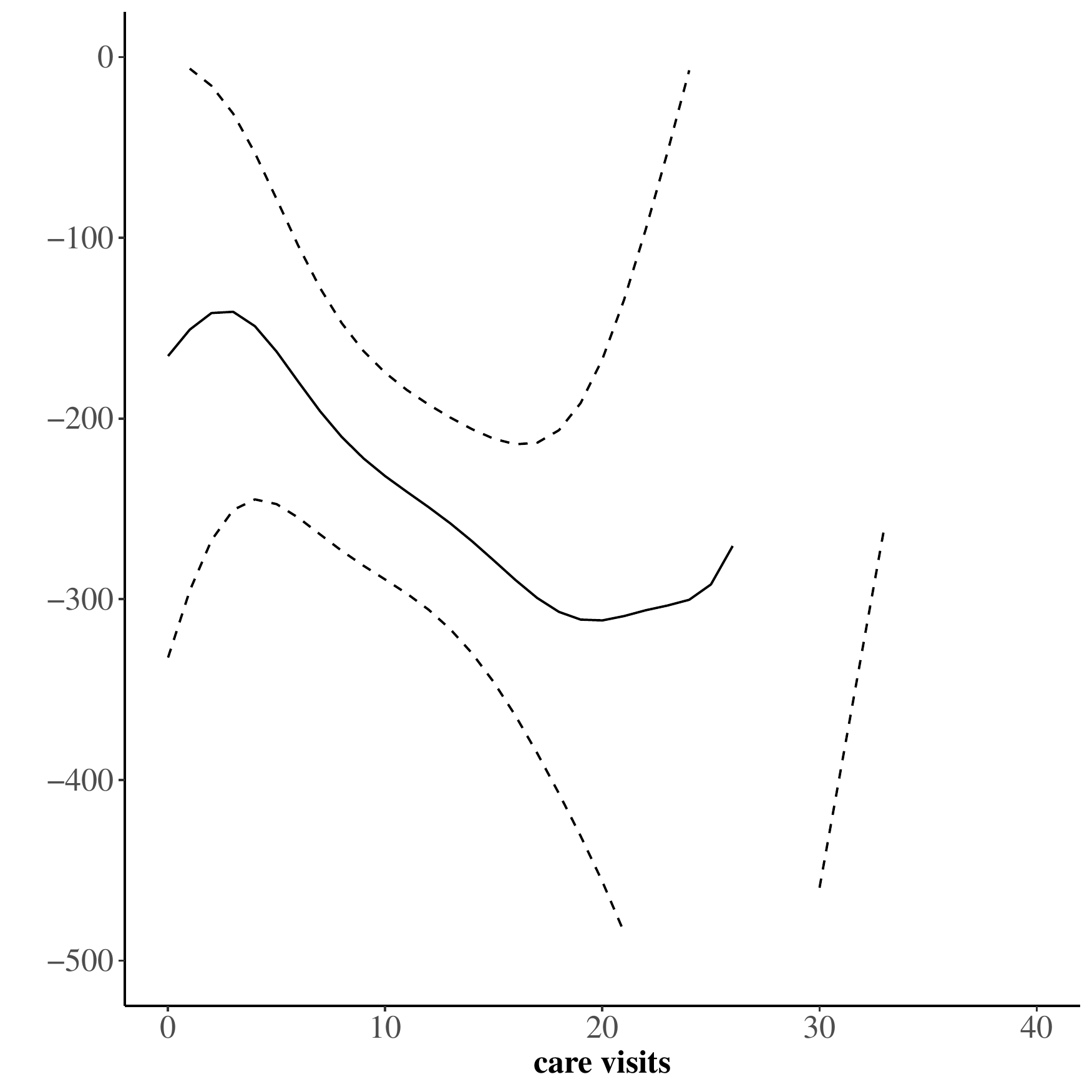}
\end{subfigure}
\begin{subfigure}{0.3\textwidth}
\caption{$0.9$ $\times$ CV choice}
\includegraphics[width=\textwidth]{figures/aipwvec_en_pc.pdf}
\end{subfigure}
\begin{subfigure}{0.3\textwidth}
\caption{$1.0$ $\times$ CV choice}
\includegraphics[width=\textwidth]{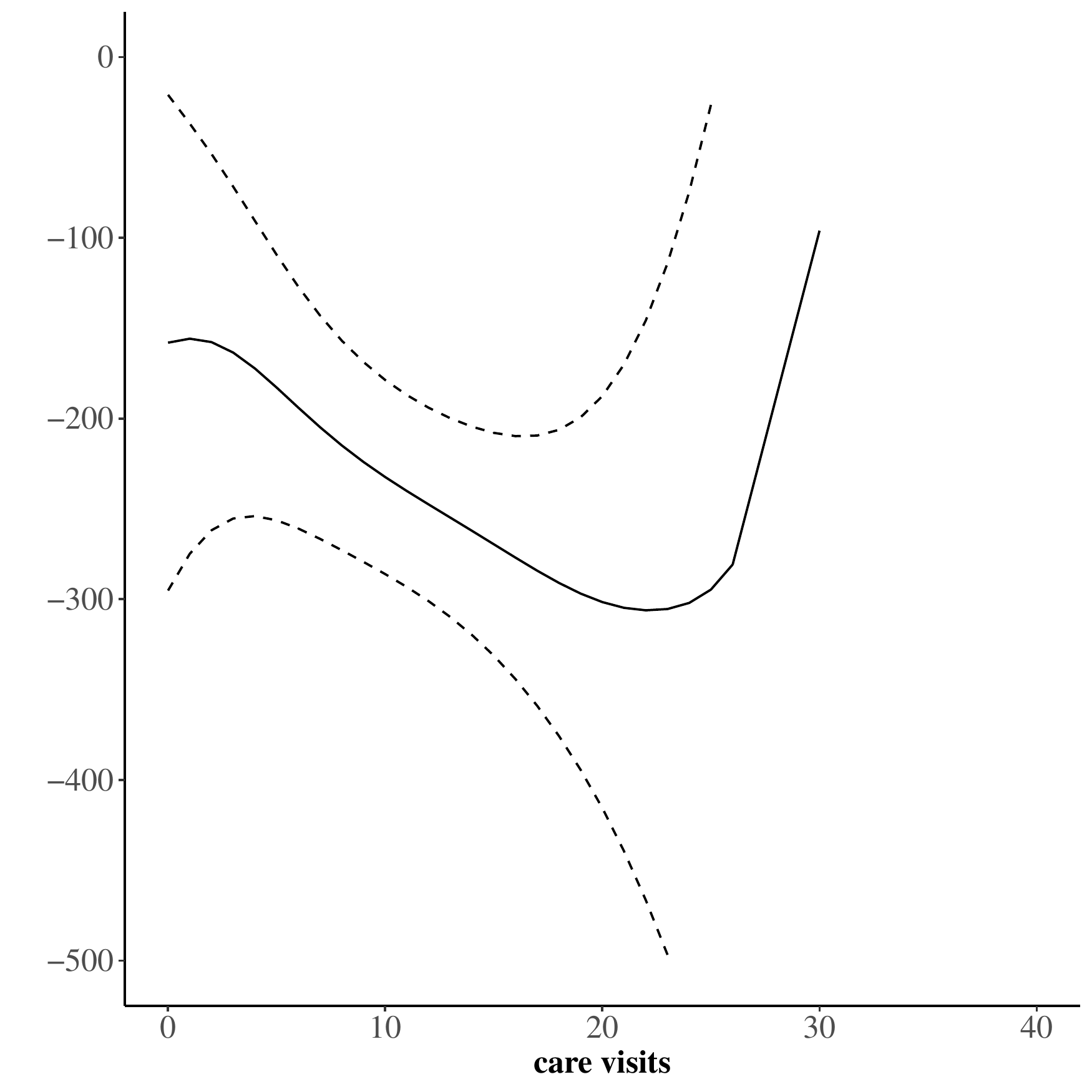}
\end{subfigure}
\begin{subfigure}{0.3\textwidth}
\caption{$1.5$ $\times$ CV choice}
\includegraphics[width=\textwidth]{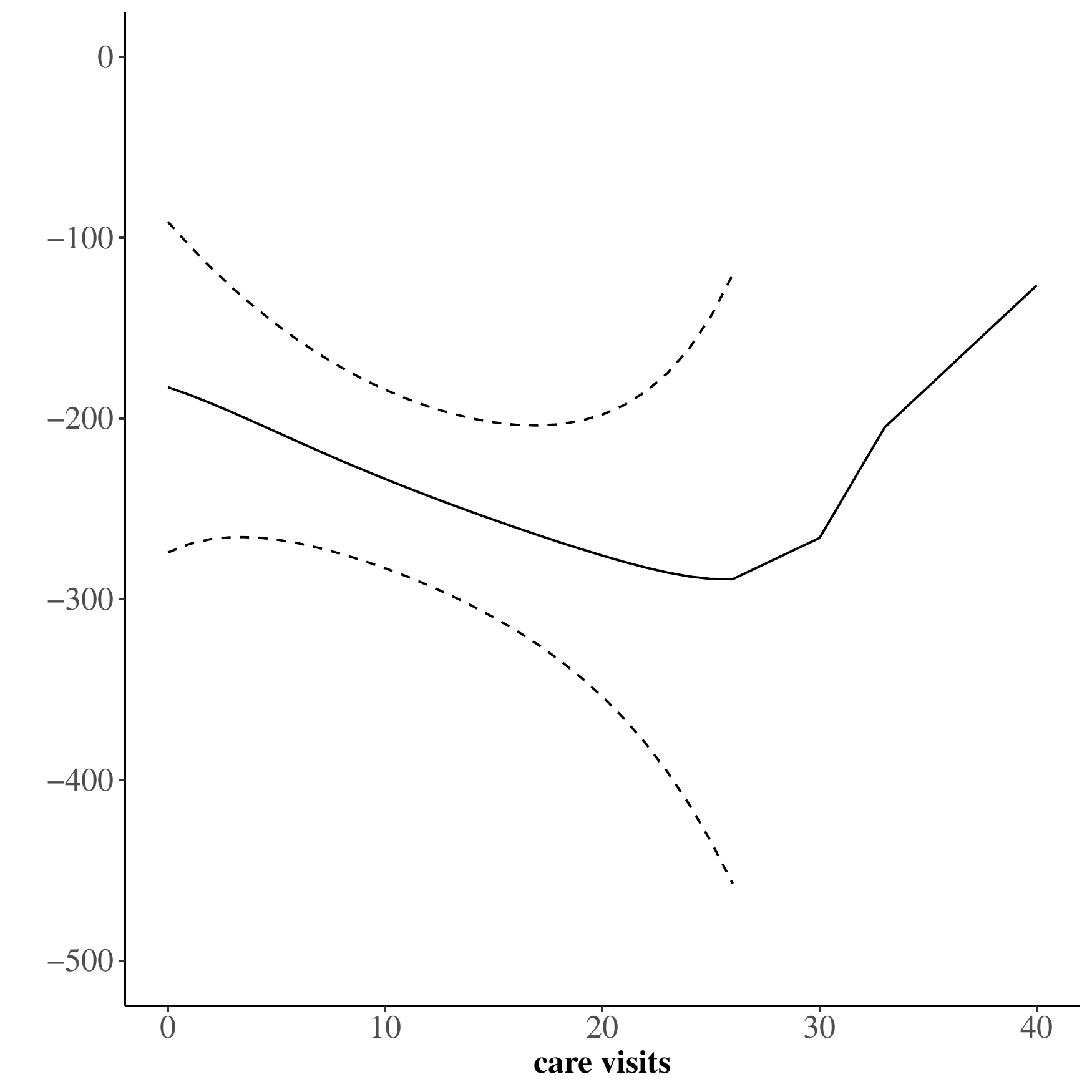}
\end{subfigure}
\label{fig:aipwbwpc}
\floatfoot{Results were obtained as described in Procedure 1 with a second-order Gaussian kernel function and different multiples of the LOOCV bandwidth choice. Nuisance parameters were estimated using an ensemble learner comprising Lasso, Elastic Net, Ridge and Random Forest. For Lasso, Ridge and Elastic Net the penalty term was chosen such that the cross-validation criterion was minimized. The ensemble weights were chosen by minimizing out-of-sample MSE. Asymptotic confidence bands are at the $95\%$ level.}
\end{figure}
\end{document}